\documentclass[aps,prd,preprint,groupedaddress,showkeys,showpacs,notitlepage]{revtex4-2}
\usepackage{graphicx,epsfig,dcolumn,bm,epic,eepic,float}
\usepackage{amsmath}
\usepackage{hyperref}
\usepackage{latexsym}
\usepackage[noabbrev]{cleveref}
\usepackage{color}
\usepackage{cancel}
\usepackage{makeidx,shortvrb,latexsym}
\begin{document}
\unitlength 1 cm
\newcommand{\nn}{\nonumber}
\newcommand{\vk}{\vec k}
\newcommand{\vp}{\vec p}
\newcommand{\vq}{\vec q}
\newcommand{\vkp}{\vec {k'}}
\newcommand{\vpp}{\vec {p'}}
\newcommand{\vqp}{\vec {q'}}
\newcommand{\bk}{{\bf k}}
\newcommand{\bp}{{\bf p}}
\newcommand{\bq}{{\bf q}}
\newcommand{\br}{{\bf r}}
\newcommand{\bR}{{\bf R}}
\newcommand{\up}{\uparrow}
\newcommand{\down}{\downarrow}
\newcommand{\fns}{\footnotesize}
\newcommand{\ns}{\normalsize}
\newcommand{\cdag}{c^{\dagger}}

\title {Three-photon productions within the $k_t$-factorization for  the ATLAS-LHC data}
\author{\textit{R. Kord Valeshabadi}}
\affiliation {Department of Physics, University of $Tehran$,
1439955961, $Tehran$, Iran.}
\author{\textit{M. Modarres} }
\altaffiliation {Corresponding author, Email:
mmodares@ut.ac.ir,Tel:+98-21-61118645, Fax:+98-21-88004781.}
\author{\textit{S. Rezaie}}
\affiliation {Department of Physics, University of $Tehran$,
1439955961, $Tehran$, Iran.}
 \date{\today}
\begin{abstract}
Recently, the ATLAS data of isolated three-photon production showed that the next-to-leading order (NLO) collinear factorization is not enough to describe experimental data. Therefore, one needs to calculate the cross section beyond the NLO, and as showed later, these data can be well described by the NNLO calculation within the collinear factorization framework. However, it is shown that the $k_t$-factorization can be quite successful in describing exclusive and high energy collision processes, henceforth we decided to calculate isolated three-photon production within this framework. In this work we use the Martin, Ryskin, and Watt unintegrated parton distribution functions (MRW UPDFs) at LO and NLO levels, in addition to parton branching (PB) UPDFs in order to calculate cross section which we utilize the KATIE parton level event generator. It will be shown that in contrast to collinear factorization, the $k_t$-factorization can describe quiet well the three-photon production ATLAS data. Interestingly our results using the NLO-MRW and PB UPDFs can cover the data within their uncertainty bands, similar to the NNLO collinear results. 
\end{abstract}
 \maketitle
\newpage
\section{Introduction}
\label{sec:I}
Precise prediction of experimental data at the LHC is one of the main challenges in high energy physics. Three isolated prompt photons in the protons-protons collision at the ATLAS \cite{three_photons_ATLAS} shows that the next-to-leading order (NLO) result is not enough to obtain a good description of the data. However, it is shown that the next-to-next-to leading order (NNLO) QCD results \cite{triple_photons_NNLO1,triple_photons_NNLO2} can nicely cover the experimental data. These predictions are based on the collinear factorization framework, and assumes that the parton enters into hard interaction is collinear to the incoming proton, i.e. $k=xP$, where $P$ is momentum of the proton, and $x$ is the fraction of the proton's momentum that the parton carries. This factorization allows us to write the hadronic cross section as a convolution of partonic cross section, $\hat{\sigma}$, and parton distribution functions (PDFs): 
\begin{equation}
\label{eq:1}
\sigma  = \sum_{i,j \in {q,g}}\int \dfrac{dx_1}{x_1}\dfrac{dx_2}{x_2}f_i(x_1,\mu^2) f_j(x_2,\mu^2)\hat{\sigma}_{ij}
\end{equation}
Where  $f_{i(j)}(x_{1(2)},\mu_F^2)$  in the above equation are the momentum weighted parton densities and related to PDFs as $f_{q(g)}(x_{1(2)},\mu_F^2)=x_{1(2)}q(g)(x_{1(2)},\mu_F^2)$. These scale dependent PDFs follow the DGLAP evolution equation \cite{DGLAP1,DGLAP2,DGLAP3} and are based on the assumption that parton enters into hard interaction emits collinear parton along the evolution ladder. Hence one has the strong ordering on the scale in a way the transverse momentum of parton is negligible with respect to the scale along the evolution ladder.

However, at large center of mass energies, $x$ becomes small and transverse momentum of parton is also comparable against the collinear component of momentum. Therefore the momentum of parton can be written as $k=xP+k_t$, where $k_t$ is transverse momentum of the parton. In contrast to collinear factorization framework, due to the important role of parton transverse momentum, no strong ordering on evolution scale exists and hence instead of the PDFs one needs transverse momentum dependent parton distribution functions (TMDs) or equivalently Unintegrated parton distribution functions (UPDFs), i.e. $f(x,k_t^2, \mu^2)$. In this framework hadronic cross section can be calculated with the help of the $k_t$-factorization, i.e.:

\begin{equation}
\label{eq:2}
\sigma  = \sum_{i,j \in {q,g}}\int \dfrac{dx_1}{x_1}\dfrac{dx_2}{dx_2} \dfrac{dk_{1,t}^2}{k_{1,t}^2}\dfrac{dk_{2,t}^2}{k_{2,t}^2} f_i(x_1, k_{1,t}^2, \mu^2) f_j(x_2, k_{2,t}^2, \mu^2)\hat{\sigma}_{ij}^*,
\end{equation}

Where $\hat{\sigma}_{ij}^*$ is the off-shell partonic cross section.

The $k_t$-factorization has shown to be successful in describing the data of non-inclusive and those belong to high energy limit observables, see reviews \cite{smallXReview1,smallXReview2,smallXReview3}. However, there are challenges in obtaining appropriate UPDFs for all partons, due to the fact that the BFKL \cite{BFKL1,BFKL2,BFKL3} and CCFM \cite{CCFM1,CCFM2,CCFM3} evolution equations were only limited to gluon. Although later CCFM UPDFs are extended to include valence quarks \cite{CCFM_valence2}, but still no complete set of CCFM UPDFs for all quark flavors exist.

One of the first approaches that defined UPDFs for all partons is named Kimber, Martin and Ryskin (KMR) \cite{KMR}. This approach utilizes the DGLAP evolution equation in a way that parton becomes $k_t$ dependent only in the last evolution step. This approach in its original form applies angular ordering not only on gluon emission but also on quark emission, too. However, later Martin, Ryskin and Watt (MRW) \cite{MRW} fix this theoretical inconsistency, by correctly applying this ordering only on gluon emission. They also extends the LO-MRW approach to the NLO level (NLO-MRW). One can find a detailed  investigation of these UPDFs in the references \cite{Mod1,Mod2,Mod3}, and also effects of using these UPDFs in cross section calculations are discussed in the references \cite{Mod16,Mod15,Mod14,Mod13,Mod12,Modphoton,single_photon_modarres,Lipatov_photon,Kord_Mod_single_jet}. 

Recently, another approach for obtaining different UPDFs introduced which solves the DGLAP evolution equation by parton branching method and collects transverse momentum of partons along the evolution ladder \cite{PB1,PB2}. This method is called parton branching (PB), and has shown remarkable success in describing experimental data \cite{PB_low_mass_drell_yann}. This method is also compared with the LO-MRW approach by calculating the Drell-Yan Z-boson $p_T^{ll}$ distribution \cite{PB_dynamical_res}, and showed that its prediction  at small and large di-lepton transverse momentum is closer to the data than the LO-MRW. 

Our goal in this work is to utilize the $k_t$-factorization to calculate three isolated prompt photon productions, which is not currently available in the literature. On the one side, we pursue to show that the $k_t$-factorization framework with only tree level Feynman diagrams at the NLO-level with the PB and NLO-MRW UPDFs can describe the data well and even comparable to the NNLO level of collinear factorization framework. On the other side, it will be shown that the NLO-MRW UPDF in the wide kinematics range is close to the PB UPDF. 

The structure of this paper is as follows: In \cref{sec:2} we give an overview of different UPDF models, including LO-MRW, NLO-MRW and PB, then we present the method of calculation and experimental cuts. In the \cref{sec:3} we discuss our results by presenting, comparing and discussing different UPDFs and also their predictions of experimental data. Finally, in the \cref{sec:4} our conclusions will be presented.

\section{Theoretical framework}
\label{sec:2}
\subsection{MRW UPDFs at LO and NLO levels}
\label{sec:2a}
LO-MRW UPDFs as introduced shortly in the introduction is based on the DGLAP evolution equation. In this approach it is assumed that parton evolves to the last evolution step collinear to the parent proton, i.e. $f_{b \in {q,g}}(x/z,k_t^2)$, where $z$ is the momentum fraction with respect to the parent parton, and also as it is obvious the scale of the DGLAP evolution is set equal to transverse momentum of parton. Then the parton has a real emission with $k_t$ comparable to the factorization scale described by  $\dfrac{\alpha_s(k_t^2)}{2\pi} P_{ab}(z)$ in the leading logarithmic approximation. Finally, the parton evolves to the factorization scale without emitting any real emission via the Sudakov form factor, i.e. $T_a(k_t^2,\mu^2)$. Therefore, the MRW UPDFs can be written as follows: 
\begin{equation}
\label{eq:3}
f_a(x,k_t^2,\mu^2) =T_a(k_t^2,\mu^2) \dfrac{\alpha_s(k_t^2)}{2\pi} \int_x^1 P_{ab}(z) f_b(\dfrac{x}{z}, k_t^2),
\end{equation}
where the Sudakov form factor is:
\begin{equation}
\label{eq:4}
T_a(k_t^2,\mu^2) = exp\bigg(-\int_{k_t^2}^{\mu^2} \dfrac{d\kappa_t^2}{\kappa_t^2}\dfrac{\alpha_s(\kappa_t^2)}{2 \pi} \sum_{b=q,g}\int_0^1 d\xi \xi P_{ba}(\xi) \bigg).
\end{equation}

One should note that the above equation is valid only in the $k_t \geq \mu_0 \sim 1 \; GeV $, and for defining UPDFs at $k_t < \mu_0$, normalization condition can be employed:
\begin{equation}
\label{eq:6}
f_a(x,\mu^2) = \int_0^{\mu^2} \dfrac{dk_t^2}{k_t^2} f_a(x,k_t^2,\mu^2).
\end{equation}
 Therefore, constraining UPDFs to satisfy the normalization condition, the following constant distribution can be obtained at $k_t < \mu_0$ \cite{wattDoubleJet, MRW}:
\begin{equation}
\label{eq:7}
\dfrac{1}{k_t^2}f_a(x,k_t^2 ,\mu^2)\Bigg\vert_{k_t < \mu_0} = \dfrac{1}{\mu_0^2}f_a(x,\mu_0^2)T_a(\mu_0^2,\mu^2)
\end{equation}

Expanding equation \ref{eq:3}, one can write the LO-MRW UPDFs in their full forms as follows:
\begin{equation}
\label{eq:8}
\begin{split}
f_q(x,k_t^2, \mu^2) = T_q(k_t^2, \mu^2) \frac{\alpha_{s}^{LO}(k_t^2)}{2\pi}\int_x^1\Big[ P_{qq}^{LO}(z)f_q^{LO}(\frac{x}{z}, k_t^2)\Theta(z_{max}-z)  \\
+ P_{qg}^{LO}(z) f_g^{LO}(\frac{x}{z}, k_t^{2})\Big]\;\mathrm{d}z,
\end{split}
\end{equation}

\begin{equation}
\label{eq:10}
\begin{split}
f_g(x,k_t^2, \mu^2) = T_g(k_t^2, \mu^2) \frac{\alpha_{s}^{LO}(k_t^2)}{2\pi}\int_x^1\Big[  P_{gg}^{LO}(z)f_g^{LO}(\frac{x}{z}, k_t^2)\Theta(z_{max}-z)\\
+ \sum_q P_{gq}^{LO}(z)f_q^{LO}(\frac{x}{z}, k_t^2)\Big]\;\mathrm{d}z,
\end{split}	
\end{equation}

with Sudakov form factors as:
\begin{equation}
\label{eq:9}
T_q(k_t^2,\mu^2) = exp\left(-\int_{k_t^2}^{\mu^2} \frac{\mathrm{d}p_t^2}{p_t^2} \frac{\alpha_{s}^{LO}(p_t^2)}{2 \pi}\int_{0}^1 P_{qq}^{LO}(\xi)\Theta(\xi_{max} - \xi)\;\mathrm{d}\xi \right),	
\end{equation}

\begin{equation}
\label{eq:11}
T_g(k_t^2,\mu^2) = exp\Big(-\int_{k_t^{2}}^{\mu^2}\frac{\mathrm{d}p_t^2}{p_t^2} \frac{\alpha_{s}^{LO}(p_t^2)}{2 \pi}\int_{0}^1 ( P_{gg}^{LO}(\xi)\Theta(\xi_{max}-\xi) \Theta(\xi - \xi_{min})
+ n_F  P_{qg}^{LO}(\xi))\;\mathrm{d}\xi \Big),
\end{equation} 
where in the \cref{eq:11}, $n_F$ is the active number of quark-antiquark flavors and $\xi_{min}=1-\xi_{max}$, and also in \cref{eq:8,eq:9,eq:10,eq:11}, $\xi_{max}$ and $z_{max}$ are introduced to avoid soft gluon emission divergence. 

The $z_{max}$ cutoff can be determined with the help of angular ordering of gluon emission in the last evolution step \cite{KMR,MRW,wattDoubleJet}, i.e:
\begin{equation}
\label{eq:12}
\mu > z \tilde{q},
\end{equation}
where $\tilde{q}$ is the rescale transverse momentum of the last step emission and is equal to  $k_t/(1-z)$. Therefore one can obtain $z_{max}$ as follows:
\begin{equation}
\label{eq:13}
\mu > z \dfrac{k_t}{(1-z)} \to z_{max} = \dfrac{\mu}{\mu + k_t}.
\end{equation}
Unitarity gives the same cutoff for $\xi_{max}$ \cite{wattDoubleJet}, i.e.:
\begin{equation}
\label{eq:14}
\xi_{max} = \dfrac{\mu}{\mu + p_t}.
\end{equation}

It is important to note that in the literature two forms of the LO-MRW formalism exist, where one is based on the Integral form discussed above and the other is based on the differential form, i.e.:

\begin{equation}
\label {eq:15}
f_a(x,k_t^2,\mu^2) = \dfrac{\partial}{\partial \log k_t^2}\big[ f_a(x,k_t^2) T_a(k_t^2,\mu^2) \big]
\end{equation}  

It is straightforward to show that the two forms are equivalent, however as shown in the reference \cite{Golec_on_KMR}, this is not correct. It is shown that using the differential form leads to discontinuous and negative behavior in the region $k_t \geq \mu$. In order to obtain same results with the integral form one needs to use the cutoff dependent PDFs, instead of the ordinary PDFs. However, as it is mentioned in the reference \cite{Golec_on_KMR}, complications which arise due to cutoff dependent PDFs can be avoided by using the integral form and ordinary PDF sets. Therefore, in this work we adopt the integral form of the LO-MRW with angular ordering constraint $z_{max}$.

The LO-MRW formalism is extended to the NLO level by the choice of the DGLAP scale as $k^2=k_t^2/(1-z)$ instead of $k_t^2$. In these UPDFs, NLO level strong coupling and splitting functions instead of LO ones are used. The consequence of using $k^2$, is an additional cutoff $\Theta(\mu^2-k^2)$ that suppresses $k^2$ to be less than $\mu^2$. 

Martin, et.al showed that using the LO splitting functions in this formalism has little effect on the UPDFs, and one can reach relatively the same UPDFs considering the splitting functions at the LO level instead of the NLO ones \cite{MRW}. Here in this work we use this simplified form and henceforth the NLO-MRW can be written as follows:

\begin{equation}
\label{eq:16}
\begin{split}
f_q(x,k_t^2, \mu^2) =\int_x^1 T_q(k^2, \mu^2) \frac{\alpha_{s}^{NLO}(k^2)}{2\pi} \Big[ P_{qq}^{LO}(z)f_q^{NLO}(\frac{x}{z}, k^2)\Theta(z_{max}-z)  \\
+ P_{qg}^{LO}(z) f_g^{NLO}(\frac{x}{z}, k^{2})\Big]\Theta(\mu^2-k^2) \;\mathrm{d}z,
\end{split}
\end{equation}

\begin{equation}
\label{eq:18}
\begin{split}
f_g(x,k_t^2, \mu^2) = \int_x^1 T_g(k^2, \mu^2) \frac{\alpha_{s}^{NLO}(k^2)}{2\pi} \Big[  P_{gg}^{LO}(z)f_g^{NLO}(\frac{x}{z}, k^2)\Theta(z_{max}-z)\\
+ \sum_q P_{gq}^{LO}(z)f_q^{NLO}(\frac{x}{z}, k^2)\Big] \Theta(\mu^2-k^2)\;\mathrm{d}z,
\end{split}	
\end{equation}

with sudakov form factors as:
\begin{equation}
\label{eq:17}
T_q(k^2,\mu^2) = exp\left(-\int_{k^2}^{\mu^2}  \frac{\mathrm{d}p^2}{p^2} \frac{\alpha_{s}^{NLO}(p^2)}{2 \pi}\; \int_{0}^{1} \;\mathrm{d}\xi \; P_{qq}^{LO}(\xi) \Theta(\xi_{max}-\xi) \right),
\end{equation}

\begin{equation}
\label{eq:19}
T_g(k^2,\mu^2) = exp\left(-\int_{k^2}^{\mu^2}  \frac{\mathrm{d}p^2}{p^2} \frac{\alpha_{s}^{NLO}(p^2)}{2 \pi}\; \int_{0}^{1} \;\mathrm{d}\xi \; \left[P_{gg}^{LO}(\xi) \Theta(\xi_{max}-\xi) \Theta(\xi-\xi_{min}) + n_F P_{qg}^{LO}(\xi)\right] \right).
\end{equation}

Here one should note that due to dependence of $k^2$ on $z$, the coupling and Sudakov form factors in \cref{eq:15,eq:17} are moved into the integral of $z$. Additionally, the cutoff $\Theta(\mu^2-k^2)$ stops the parton to have momentum larger than the factorization scale.  

Before finishing this section an important point is in order here. The LO-MRW in its original form has no dimension, while, some other UPDF sets in the literature have $1/(GeV^2)$ dimension. This results in different hadronic cross section formula with $1/(k_t^2)$ in the denominator of the \cref{eq:2} to be moved into the UPDFs, and hence one has $F(x,k_t^2, \mu^2) = \dfrac{f(x,k_t^2, \mu^2)}{k_t^2}$, i.e.:
\begin{equation}
\label{eq:20}
\sigma  = \sum_{i,j \in {q,g}}\int \dfrac{dx_1}{x_1}\dfrac{dx_2}{dx_2} dk_{1,t}^2 dk_{2,t}^2 F_i(x_1, k_{1,t}^2, \mu^2) F_j(x_2, k_{2,t}^2, \mu^2)\hat{\sigma}_{ij}.
\end{equation} 

In addition to this change, it is straightforward that the normalization condition changes as follows:
\begin{equation}
\label{eq:21}
f_a(x,\mu^2) = \int_0^{\mu^2} dk_t^2 F_a(x,k_t^2,\mu^2),
\end{equation}

\subsection{PB UPDFs}
\label{sec:2b}
Another approach which allows us to obtain UPDFs for both quark and gluon is called parton branching UPDFs. This method allows to obtain UPDFs by solving the DGLAP evolution equation iteratively with Monte Carlo method and by calculating transverse momentum at every splitting kernel. This method imposes angular ordering in addition to virtuality ordering along the evolution ladder. UPDFs distributions at the initial scale $\mu_0$ is chosen to have a factorized form of dependent on the transverse momentum via Gaussian distribution and a parameterized form with dependency on $x$ and $\mu_0^2$ \cite{PB_set1_2,Pegasus}. Generally parton branching UPDFs can be written as follows:
\begin{equation}
\label{eq:22}
\begin{split}
F_a(x, k_t^2,\mu^2) = \Delta_a(\mu^2) F_a(x, k_t^2, \mu_0^2) + \sum_b \int \dfrac{d^2\bm{q}{^\prime}}{\pi \bm{q}^{\prime 2}} \dfrac{\Delta_a(\mu^2)}{\Delta_a(\bm{q}^{\prime 2})} \Theta(\mu^2-\bm{q^{\prime 2}} )  \Theta(\bm{q}^{\prime 2} - \mu_0^2) \\
\int_x^{z_M} \dfrac{dz}{z} P_{ab}^{R}(\alpha_s,z) F_b(\dfrac{x}{z}, k^{\prime 2}_t, \bm{q^{\prime 2}}).
\end{split}
\end{equation}

Where $\bm{k}^{\prime}_t= \bm{q^\prime}(1-z) + \bm{k}_t$ and the $P_{ab}^{R}(\alpha_s,z)$ is the splitting function separated into two parts, one contains soft gluon emission singularity and the other contains logarithmic and analytic terms, see reference \cite{PB_set1_2}. Additionally, 
the Sudakov form factor $\Delta$ is as follows:
\begin{equation}
\label{eq:23}
\Delta_a(\mu^2) = exp\bigg(-\sum_b \int_{\mu_0^2}^{\mu^2} \dfrac{d\mu^{\prime^2}}{\mu^{\prime 2}} \int_0^{z_M} dz z P_{ba}^{R}(\alpha_s,z)\bigg),
\end{equation}
where $z_M$ is the soft gluon resolution scale and separates real and no-real emissions.

PB UPDFs can be generated by employing TMDlib \cite{TMDLIB} which provides different UPDF sets. In this work we utilize PB-NLO-HERAI+II-2018-set2 (PB18-set2) \cite{PB_set1_2} UPDF set  which are obtained by fitting to the experimental data of HERA I+II. In this set the $z_M$ is fixed to $0.99999$, but as it is pointed out in \cite{PB_set1_2,PB2} due to using angular ordering, UPDFs are stable with respect to variations of $z_M$. Another important point about PB18-set2 is that the emitted parton transverse momentum is used as the argument of the coupling constant, while for the other PB set PB-NLO-HERAI+II-2018-set1 (PB18-set1) available in the TMDlib the evolution scale $\mu$ is used. It is shown in \cite{PB_set1_2} that these two sets give similar results for large $k_t$, however the difference is mostly shown itself at small $k_t$ in a way that $q_T^{ll}$ of Z Boson Drell-Yann spectrum at small dilepton transverse momentum can be better described by PB18-set2. This is the reason we stick to PB18-set2 for calculating differential cross sections of three photons production, and for simplicity we adopt the PB name alone in what follows rather than PB18-set2.        

\subsection{Method of Calculation and Experimenta Cuts}
\label{sec:2c}
We calculate three-photon productions in proton-proton collisions at center of mass energy of $8\; TeV$ within the $k_t$-factorization framework in accordance with the corresponding ATLAS experimental data \cite{three_photons_ATLAS}. The calculation is performed with the KATIE \cite{KATIE} parton-level event generator with $n_F=5$. In the following, we first give a brief review of the KATIE and then present the cut requirements of the ATLAS three-photon production \cite{three_photons_ATLAS} experiment.   

The KATIE is the parton-level event generator which can produce parton-level events at tree level for various number of final state particles with off-shell, in addition to on-shell kinematics. This generator can be either linked to TMDlib and use different UPDFs via this library or one can produce grid files which then read by the KATIE for event generation. Here, we adopt the first method for using PB UPDFs and the latter method for using the LO-MRW and NLO-MRW UPDFs in our cross section calculation. It should be noted that the NLO-MRW UPDFs and the integral form of the LO-MRW UPDFs are not available in the TMDlib yet. Unfortunately, the LO-MRW grid files with the name MRW-CT10nlo \cite{MRW-CT10NLO} in TMDlib are based on the differential form of the LO-MRW and as discussed in the section \ref{sec:2a}, they can be problematic for cross section calculation. Therefore, we generate grids files of the LO and NLO-MRW UPDFs for each parton and also provide them publicly \footnote{LO and NLO-MRW grid files can be download from this \href{https://drive.google.com/file/d/1K5fjkVfffHTtJ33E08dmg1EXx9TH5NOV/view?usp=sharing}{link}} . For the LO and NLO-MRW input PDFs we utilize MMHT2014 PDFs set \cite{MMHT}. These grid files as mentioned in the \cite{KATIE} must be comprised of four columns according to $ln(x)$, $ln(k_t^2)$, $ln(\mu^2)$ and $F(x, k_t^2, \mu^2)$. 

In calculating cross section two sub-processes $q + \overline{q} \to \gamma \gamma \gamma$ and $q + g \to \gamma \gamma \gamma + q$ are considered, and also the following experimental cuts are imposed on our calculation:
\begin{enumerate}
	\item Photons need to be separated from each other by $\Delta R_{ij} > 0.45$, where $\Delta R_{ij}=\sqrt{(\eta_i - \eta_j)^2+(\phi_i-\phi_j)^2}$.
	\item Transverse energy of the three photons with highest transverse energies are:  $E_T^{\gamma_1} > 27\; GeV$,  $E_T^{\gamma_2} > 22\; GeV$, $E_T^{\gamma_3} > 15 \; GeV$, where $E_T^{\gamma_1}$, $E_T^{\gamma_2}$ and $E_T^{\gamma_3}$ are the photons with the highest, second highest and softest transverse energies.
	\item All photons must have pseudo-rapidities $0 \leq |\eta^{\gamma}| \leq 1.37$ or $1.56 \leq |\eta^{\gamma}| \leq 2.37$.
	\item Three photon invariant mass, $m^{\gamma_1 \gamma_2 \gamma_3}$, is larger than $50 \; GeV$.  
	\item Instead of the standard isolation cone implemented in the experiment, we impose smooth isolation cone \cite{smooth_cone} with benefits that on the one side it regularizes photon collinear divergence and on the other side it suppresses fragmentation contribution \cite{three_photons_ATLAS}.
	
	The smooth isolated cone enforces the transverse energies of particles around the distance $\Delta R \leq  R_0$ from each photon to be:
	\begin{equation}
	E_T^{iso}(\Delta R) < E_T^{max} \dfrac{1- \cos{\Delta R}}{1-\cos{R_0}},
	\end{equation}   
	where $E_T^{iso}$ is the sum of transverse energies of the particles around the photon in the distance less than $\Delta R$. In this work, our calculation is limited to the parton level and therefore we have only one particle in the final state. Additionally similar to the reference \cite{triple_photons_NNLO2}, we set $R_0 = 0.4$ and $E_T^{max} = 10 \; GeV$.
\end{enumerate}

We choose $\mu_{F,R}^{central}=\sqrt{p^2_{\gamma \gamma \gamma, T} + m_{\gamma \gamma \gamma}^2}$ as the factorization and renormalization scales. Additionally, to estimate the scale uncertainty of our calculation we repeat the same event generation one with $\mu_{F,R}^{upper} = 2 \mu_{F,R}^{central}$ and the other with $\mu_{F,R}^{lower} = 0.5 \mu_{F,R}^{central}$.

\section{Results and Discussions}
 \label{sec:3}
Before presenting the results, it is important to gain an insight of different UPDF models to understand the difference and similarity of them in cross section predictions.

\begin{figure}
	\includegraphics[width=15cm, height=9cm]{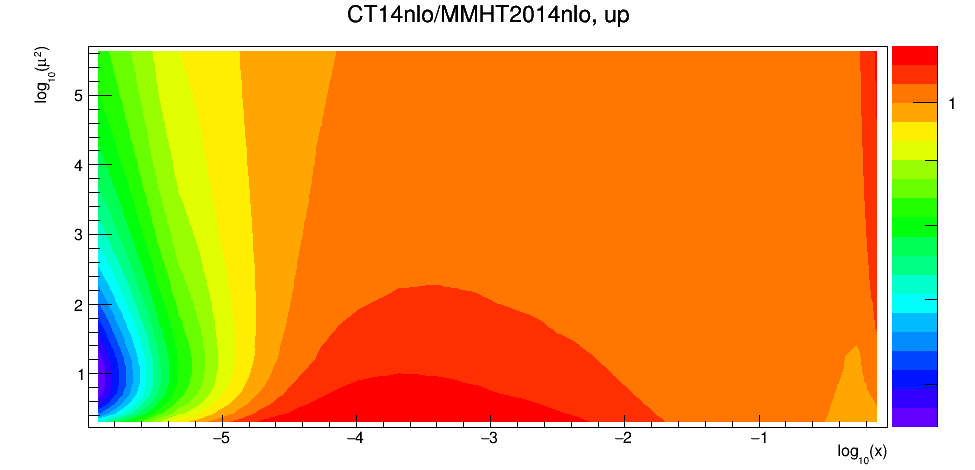}
	\includegraphics[width=15cm, height=9cm]{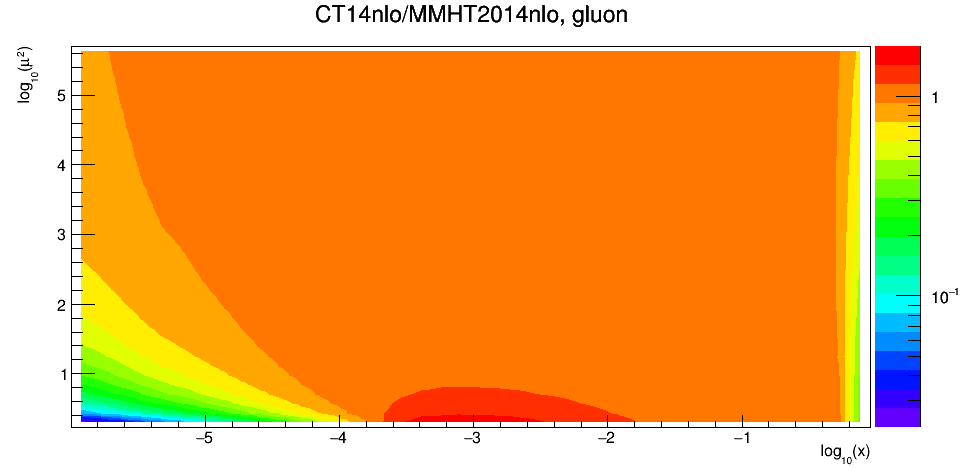}
	\caption
	{ The top (bottom) panel shows the fraction of CT14nlo/MMHT2014nlo for up quark(gluon) respectively.	
	}
	\label{fig:1}
\end{figure}

One of the most important elements of the LO and NLO-MRW UPDFs is the input PDFs. These PDFs are provided by different theoretical groups for instance MMHT2014 \cite{MMHT}, CT14 \cite{CT14nlo}, NNPDF \cite{NNPDF}. The difference between PDFs sets are mostly due to the choice of heavy quark treatment and the data sets they use for the initial scale fitting \cite{PB_set1_2}. In order to see how our results may be affected with the different input PDF sets choices, we as an example, analyze our adopted PDF set, i.e. MMHT2014, with another famous PDF set, i.e. CT14 at the NLO level by plotting the fraction of them, i.e. $CT14nlo/MMHT2014nlo$, for up and gluon PDFs in $\log_{10}x-\log_{10}\mu^2$ space. For presenting such a comparison and using these two PDF sets, LHAPDF library  \cite{LHAPDF6}  is adopted. It can be seen in the \cref{fig:1} that these two PDF sets are in accordance with each other in most $\log_{10}x-\log_{10}\mu^2$ space. Although, a relatively significant difference between gluon PDF of $CT14nlo$ and $MMHT2014nlo$ at large $log_{10}(x) \geq 0.5 $ and small $log_{10}(x) \leq -4.2$ is observed which does not play any significant role in our calculation for the energy range of three photon production experiment. To generate our results, we use PDF sets of $\textrm{MMHT2014nlo68cl}$ and $\textrm{MMHT2014lo68cl}$ for the LO-MRW and NLO-MRW UPDFs with the help of LHAPDF \cite{LHAPDF6} interface library.

\begin{figure}
	\includegraphics[width=8cm, height=9cm]{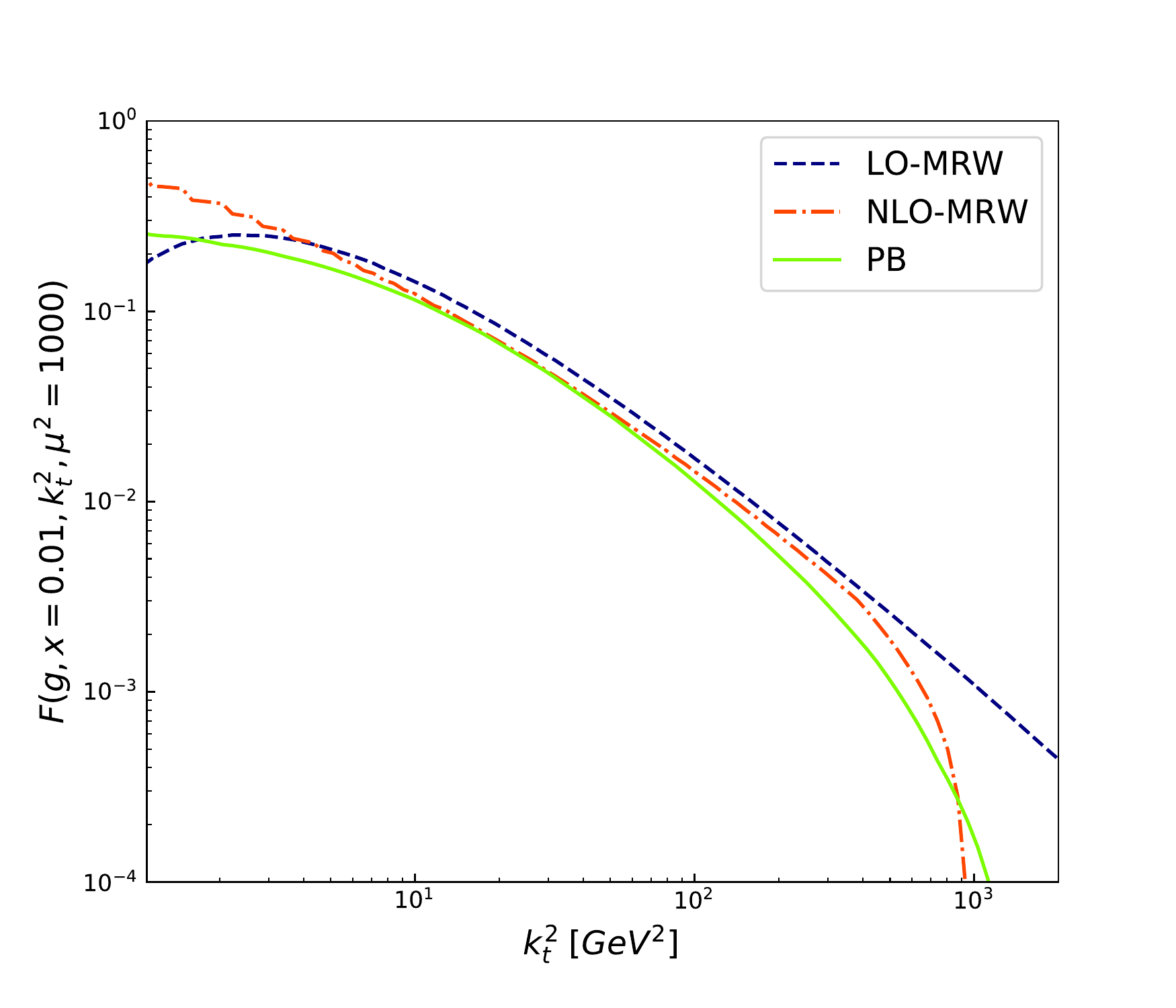}
	\includegraphics[width=8cm, height=9cm]{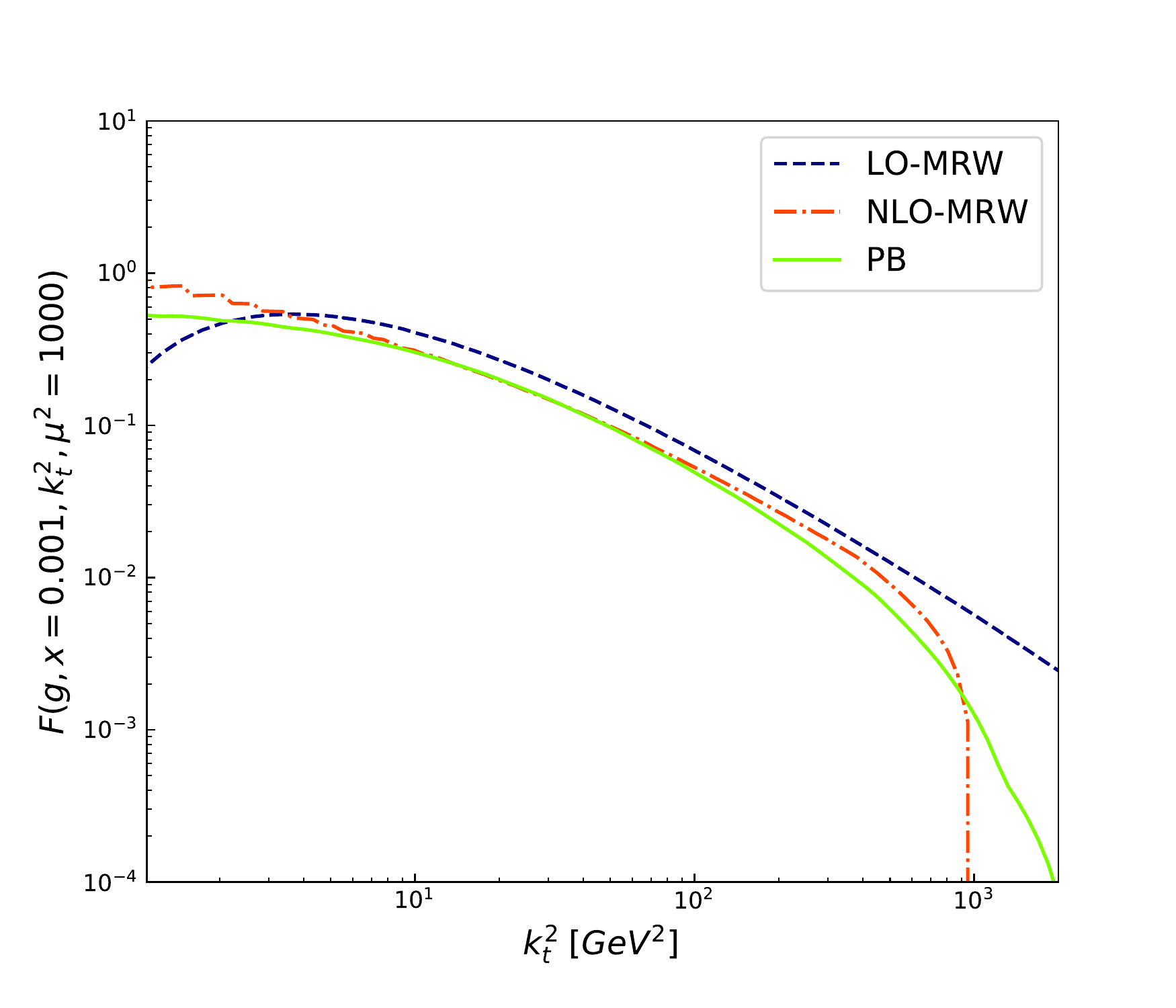}
	\includegraphics[width=8cm, height=9cm]{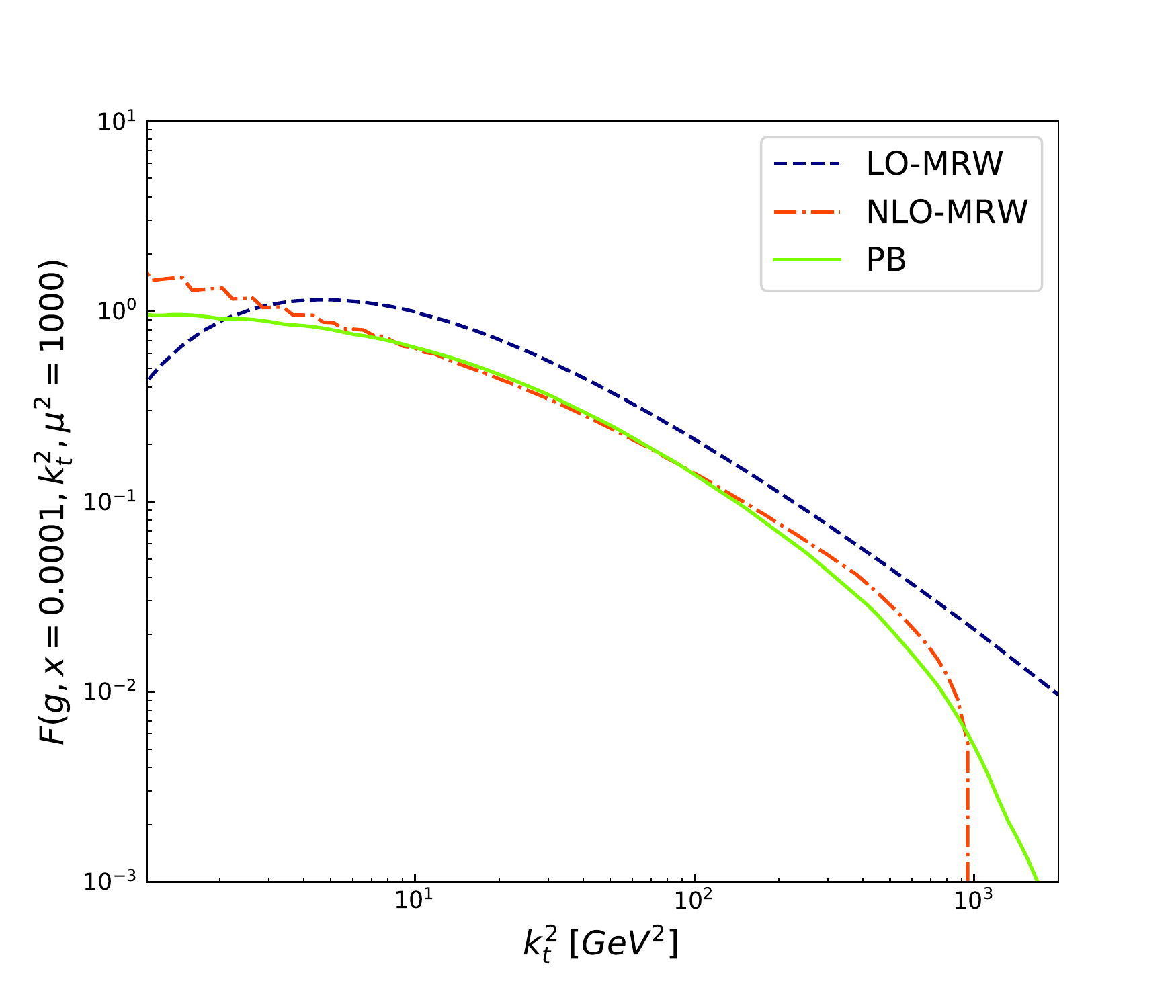}
	\caption
	{ The plots of LO-MRW, NLO-MRW and PB gluon UPDFs with respect to $k_t^2$ for different $x$ values as denoted in each plot are shown.	
	}
	
	\label{fig:2}
\end{figure}

\begin{figure}
	\includegraphics[width=8cm, height=9cm]{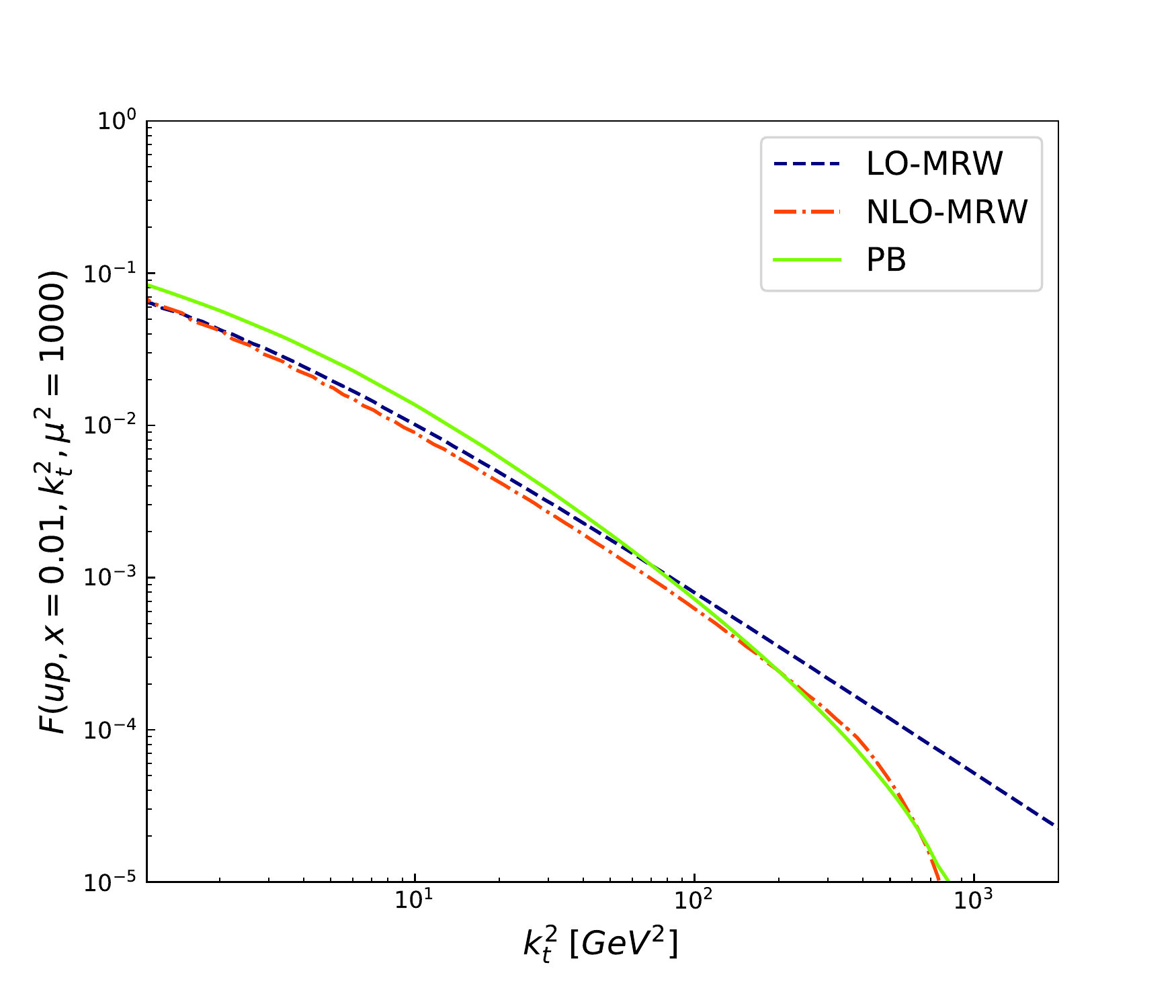}
	\includegraphics[width=8cm, height=9cm]{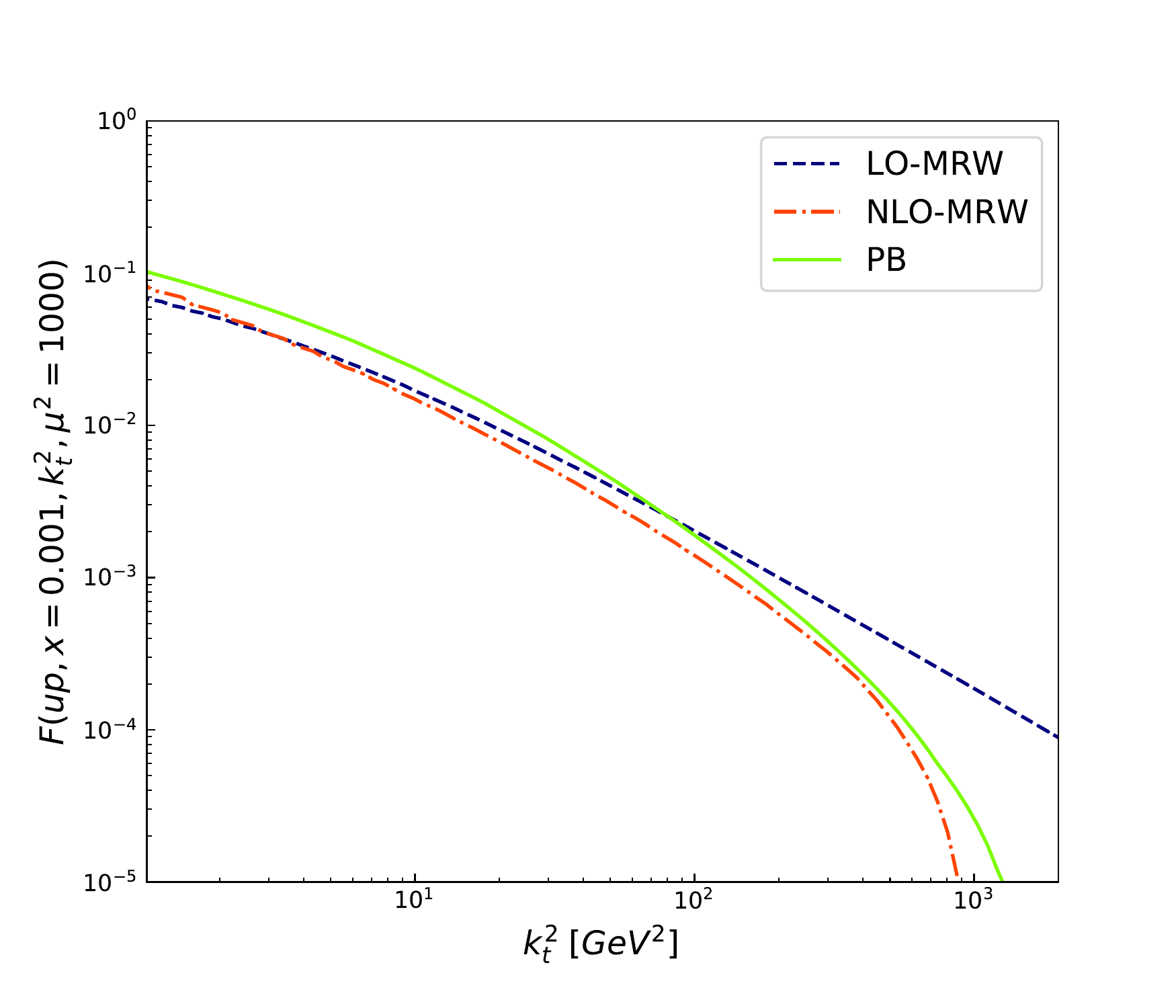}
	\includegraphics[width=8cm, height=9cm]{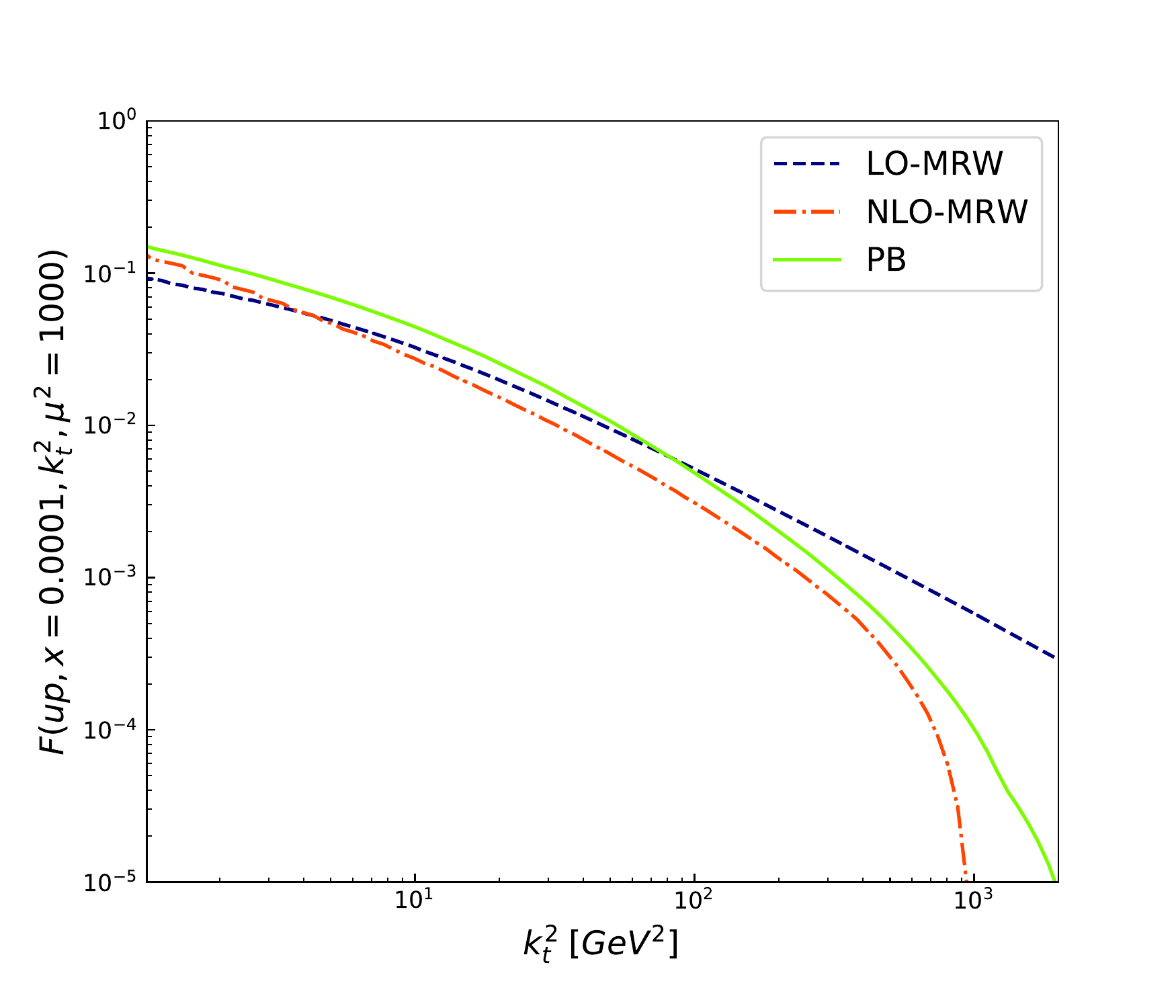}
	\caption
	{ The same as \cref{fig:2} but for up quark.	
	}
	\label{fig:3}
\end{figure}

In the \cref{fig:2,fig:3}, a comparison between different UPDF models for up quark and gluon at $\mu^2=1000$ and for $x=0.0001$, $x=0.001$ and $x=0.01$ is shown. As can be seen in these two figures, UPDFs of the NLO-MRW and PB are larger than the LO-MRW UPDFs at small transverse momentum, while they become smaller at large transverse momentum. The reason for steep decrease of NLO-MRW UPDFs at large transverse momentum is due to the cutoff $\Theta(\mu^2-k^2)$, where it constrains the transverse momentum to the region less than the factorization scale. While, because the LO-MRW UPDFs have no cutoff on the non-diagonal terms of the DGLAP splitting function, this leads to LO-MRW UPDFs becomes much larger at $k_t$ close to $\mu$ with respect to other two UPDF models. Finally, it can be seen that while gluon UPDF models at middle transverse momentum are relatively similar to each other, for up quark the PB UPDF becomes larger with respect to other two UPDF models.

\begin{figure}
	\includegraphics[width=15cm, height=9cm]{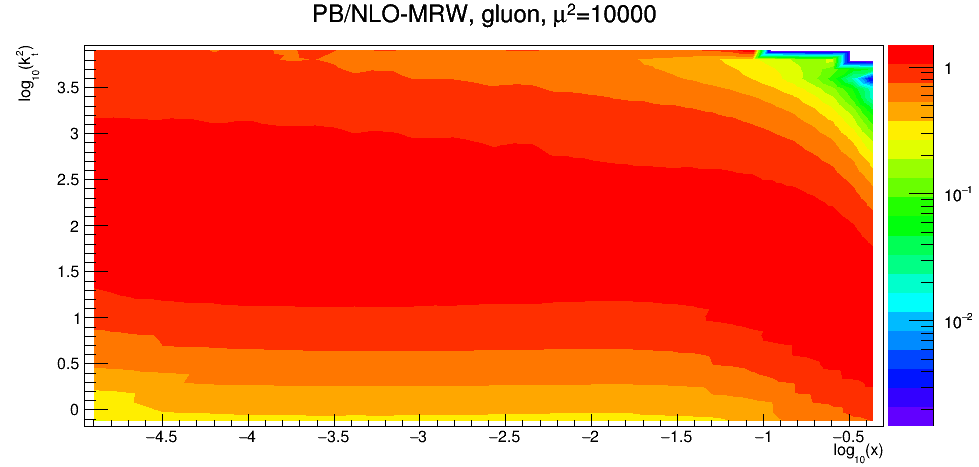}
	\includegraphics[width=15cm, height=9cm]{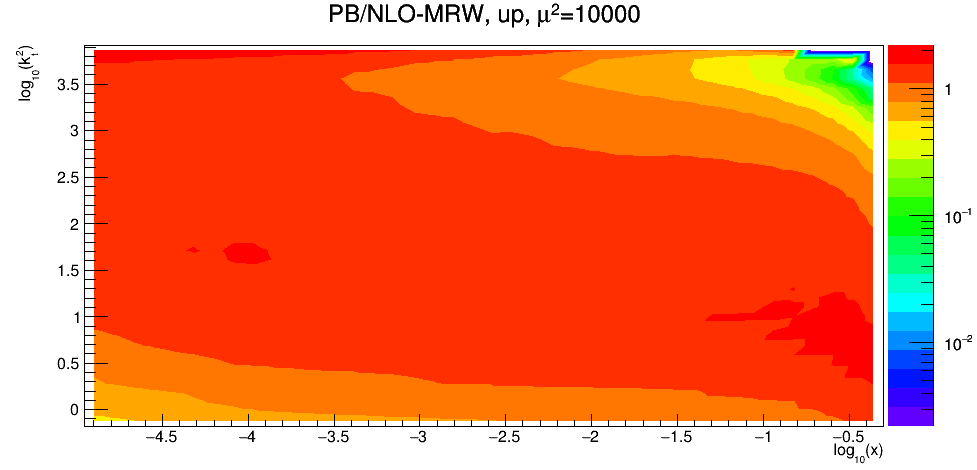}
	\caption
	{ The top (bottom) panel shows the fraction of $NLO-MRW/PB$ for gluon (up) quark at $\mu^2=10000$, respectively.	
	}
	\label{fig:4}
\end{figure}

As can be seen in the  \cref{fig:4}, the degree of similarity between two UPDF models of NLO-MRW and PB is shown by plotting the fraction of NLO-MRW/PB for gluon and up quark in $\log_{10}x-\log_{10}k_t^2$ space at $\mu^2=10000$. These plots enable us to gain a better insight of each UPDF models by covering a wide range of $x$ and $k_t^2$ region. As it is shown in in this figure, the gluon UPDFs of the NLO-MRW and PB are in close agreement to each other in most region. While for the up quark UPDFs, the NLO-MRW only similar at small and large $k_t$. As mentioned before the NLO-MRW becomes smaller than the PB up UPDFs at middle transverse momentum.

\begin{figure}
	\includegraphics[width=15cm, height=9cm]{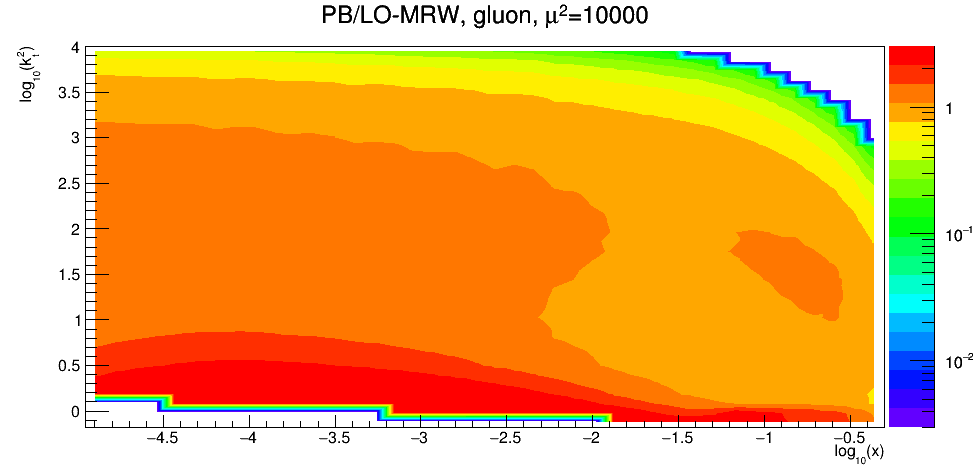}
	\includegraphics[width=15cm, height=9cm]{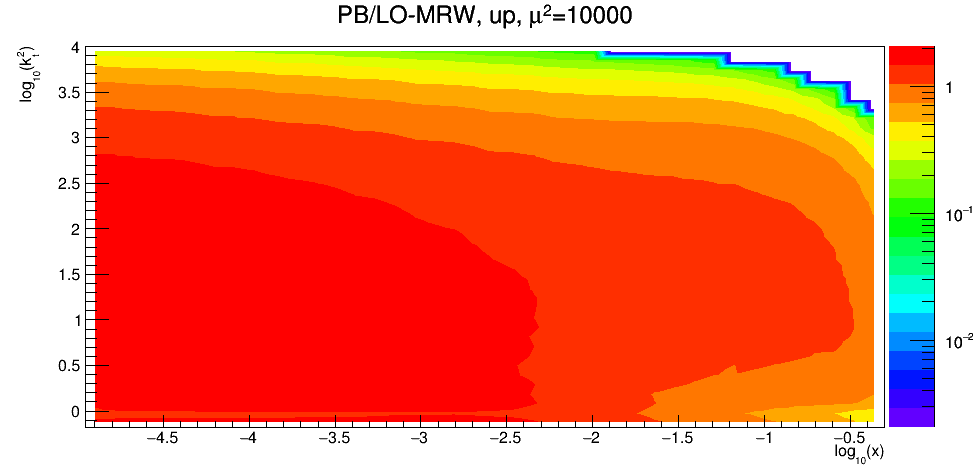}
	\caption
	{ The top (bottom) panel shows the fraction of $PB/LO-MRW$ for gluon (up) quark at $\mu^2=10000$, respectively.	
	}
	\label{fig:5}
\end{figure}

In the \cref{fig:5} the degree of similarity between the UPDFs of LO-MRW and PB is investigated with the help of the fraction PB/LO-MRW at $\mu^2=10000$ in $\log_{10}x-\log_{10}k_t^2$ space. A cut is also imposed to remove regions with large difference. As can be seen in this figure, the same behavior as before is observed where gluon UPDFs of the LO-MRW is smaller than the PB in the small $k_t$, and relatively similar to each other at middle $k_t$, while LO-MRW becomes larger than PB in $k_t \sim \mu$. For the MRW up UPDFs, this similarity with PB becomes worse compared to the similarity between PB and NLO-MRW up UPDFs. The up UPDF of the LO-MRW and PB is only close to each other in the limited $\log_{10}x-\log_{10}k_t^2$ region.

As observed in the \cref{fig:2,fig:3,fig:4,fig:5}, UPDFs of the NLO-MRW and PB are more similar to each other than the LO-MRW and PB ones. Therefore, one could expect their predictions of experimental results also be close to each other. Now, we seek to investigate these UPDFs and also the $k_t$-factorization framework by means of comparing the differential cross section predictions of UPDF models with each other and also the NNLO \cite{triple_photons_NNLO2} results of the collinear factorization framework. In the reference \cite{triple_photons_NNLO2}, cross section is calculated for different choices of the factorization/renormalization scales, however the chosen central scale choice is $\mu_{F,R} = \dfrac{1}{4}\sum_{i=1}^{3} E_{T}^{\gamma_i}$. We use the data in ancillary files provided along with the manuscript of the reference \cite{triple_photons_NNLO2} to show comparison between the NNLO collinear with our $k_t$-factorization results. 

In \cref{table:1}, a comparison of fiducial cross sections of different UPDF models and also the NNLO collinear results are presented. It is interesting to note that the cross section prediction of all our UPDF models cover the experimental cross section $\sigma^{expriment} = 72.6\pm 6.5 \;(stat) \pm 9.2 (sys)\; fb$, where stat and sys denote statistical and systematic uncertainties. It can be seen that the NLO-MRW and PB UPDF models central value cross section predictions are more in accordance with the experiment with respect to the LO-MRW UPDF prediction, where this model tends to overestimate the experimental cross section. The reason for such behavior is the large role of LO-MRW UPDFs at $k_t \sim \mu$ with respect to the NLO-MRW and PB UPDF models. It should be noted that large tail of the LO-MRW UPDF models with respect to the PB is investigated in the reference \cite{PB_dynamical_res}. The key point is that one cannot see the unwanted behavior of the LO-MRW UPDF model at large parton transverse momentum by imposing virtuality ordering in the NLO-MRW  UPDF model, and as a result of this constraint, the NLO-MRW has better performance with respect to the LO-MRW in prediction of the fiducial cross section.

 \begin{table}[h]
	\centering
	\begin{tabular}[b]
		{|c | c| c| }
		\hline
		 & $\sigma^{prediction} $  \\
		\hline
		LO-MRW     & $86.17^{+8.31}_{-2.64}\;fb$ \\
		\hline
		NLO-MRW	  & $64.66^{+12.54}_{-14.69} \;fb$    \\
		\hline
		PB  	  & $68.99^{+8.66}_{-9.51}\;fb$   \\
		\hline  
		NNLO Collinear  & $67.46^{+7.39}_{-4.91}\;fb$   \\
		\hline 
	\end{tabular}
	\caption{Predictions of different UPDFs models an the NNLO collinear for the inclusive fiducial of three-photon cross section. The upper and lower limit of the cross section is due to scale uncertainty.}
	\label{table:1}
\end{table}

\begin{figure}
	\includegraphics[width=8cm, height=9cm]{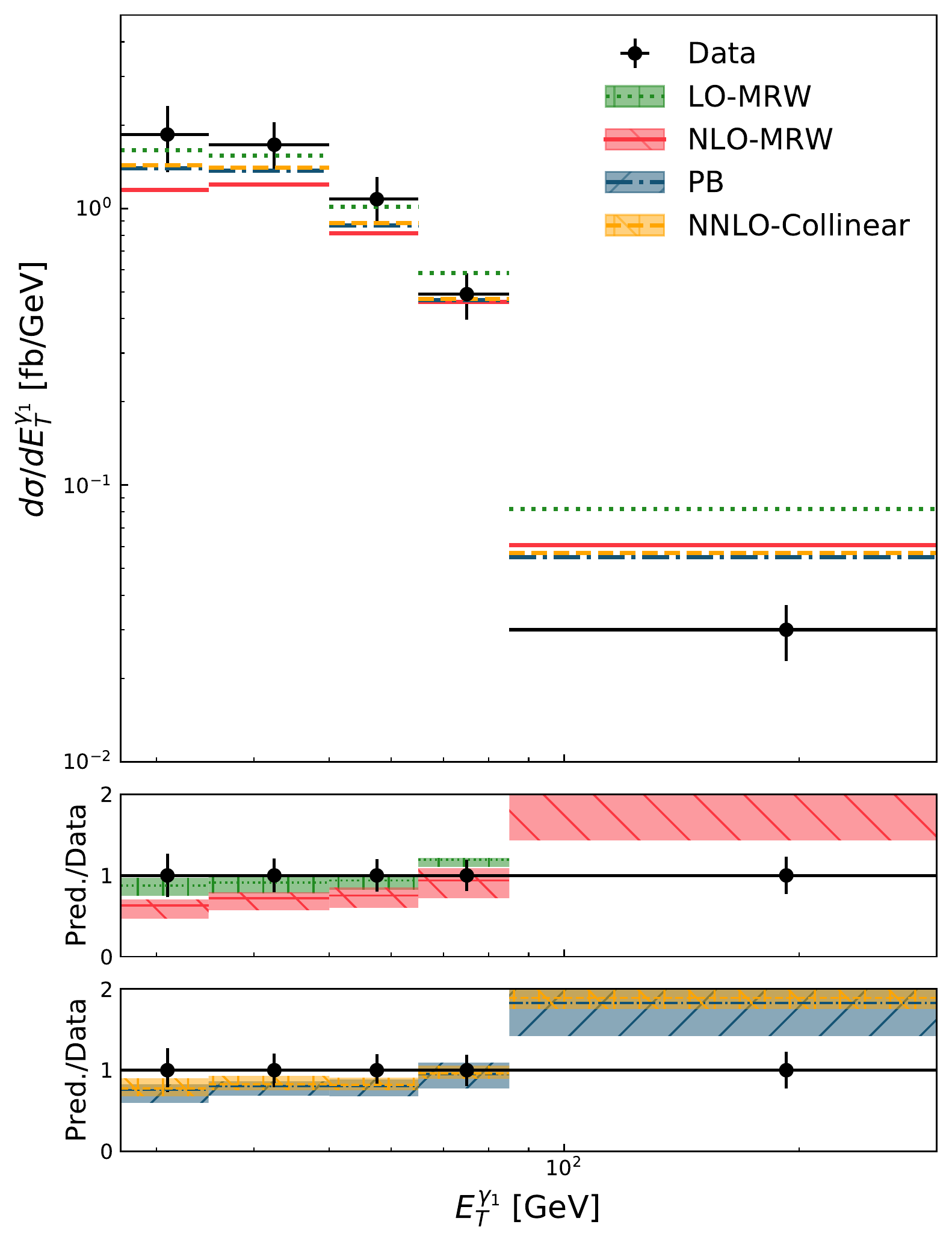}
	\includegraphics[width=8cm, height=9cm]{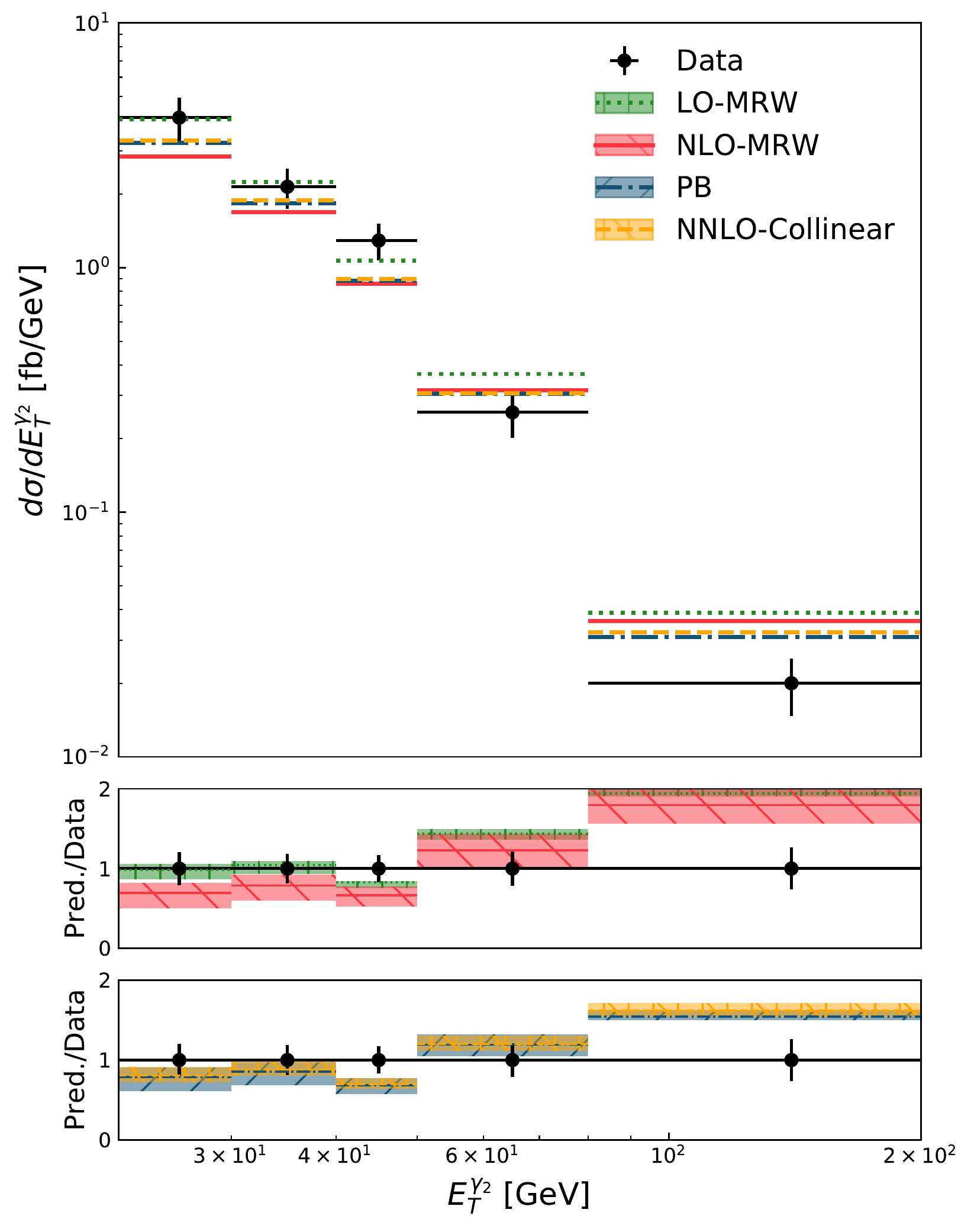}
	\includegraphics[width=8cm, height=9cm]{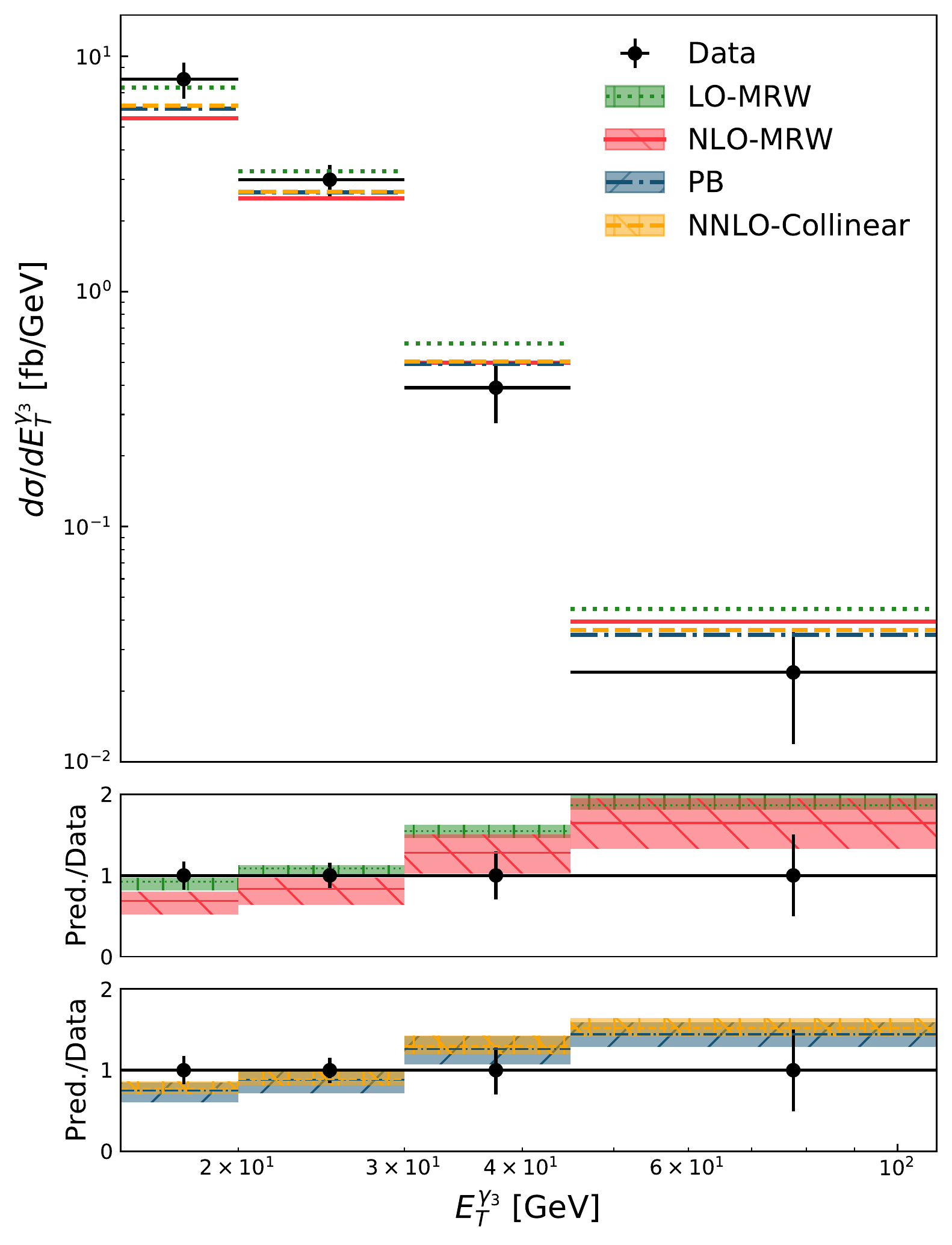}
	\caption
	{ In the top panel of each plot the differential cross sections for different UPDF models and also NNLO collinear with respect to $E_T^{\gamma_1}$, $E_T^{\gamma_2}$ and $E_T^{\gamma_3}$ are shown and compared to the ATLAS experimental data \cite{three_photons_ATLAS} as denoted in each figure. Additionally, in the two bottom panels in each plot, prediction/data of each UPDF models and NNLO collinear with their scale uncertainties are shown.	
	}
	\label{fig:6}
\end{figure}

In the \cref{fig:6}, differential cross sections predictions with respect to the transverse energies of the hardest, second hardest and the softest final state photons, i.e. $E_T^{\gamma_1}$, $E_T^{\gamma_2}$ and $E_T^{\gamma_3}$ are shown. As could be expected, one can see that the LO-MRW prediction is slightly larger than the other two UPDFs results. The interesting point here is that the NLO-MRW and PB UPDFs are more in agreement with each other and also are covering the results within their uncertainty band. Above all, it can be surprising to see that the $k_t$-factorization prediction with the PB UPDFs and the NLO-MRW UPDFs are even close to the results of the NNLO collinear factorization \cite{triple_photons_NNLO2}. While, as can be seen in the reference \cite{three_photons_ATLAS}, the NLO collinear results undershoot the data.     

\begin{figure}
	\includegraphics[width=8cm, height=9cm]{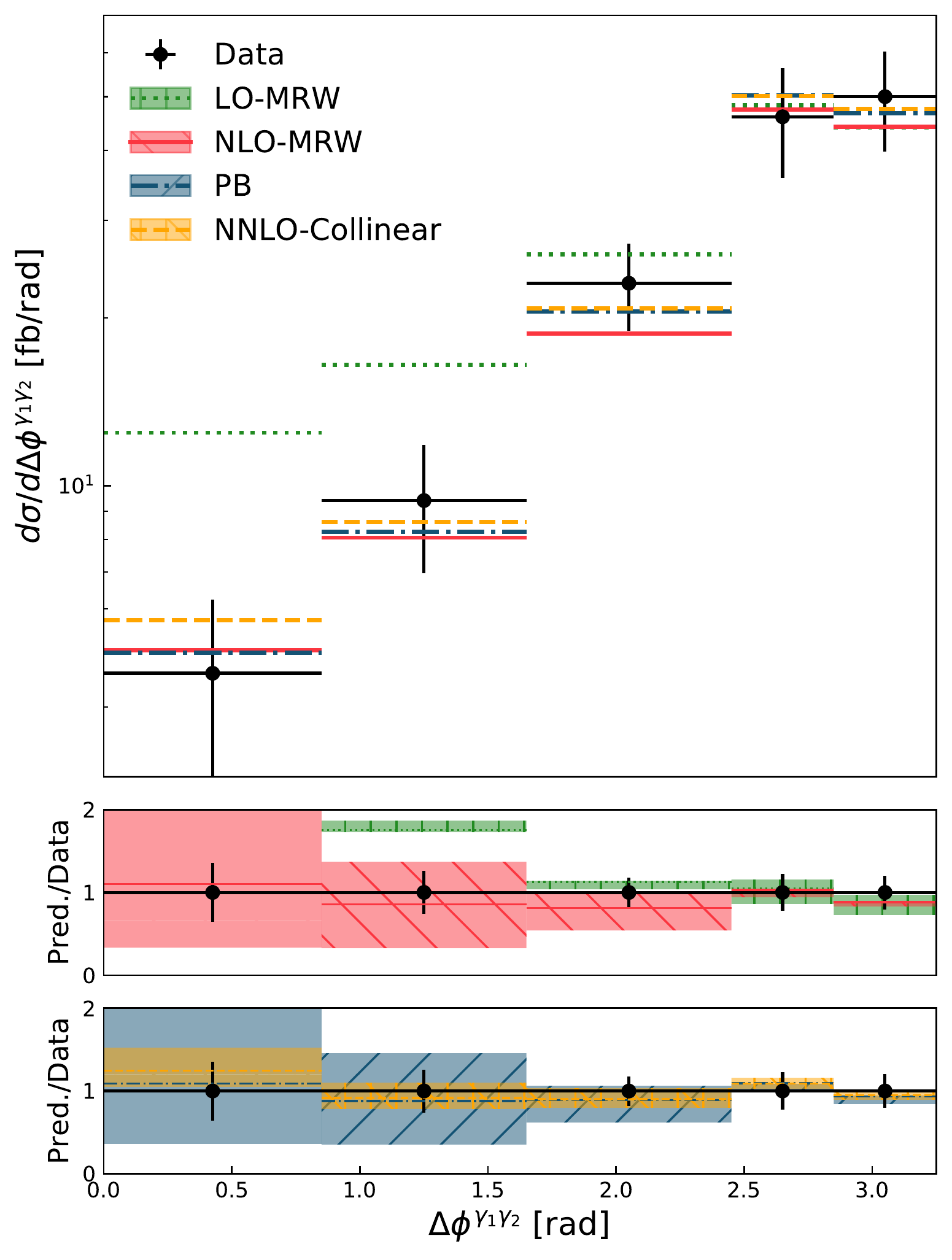}
	\includegraphics[width=8cm, height=9cm]{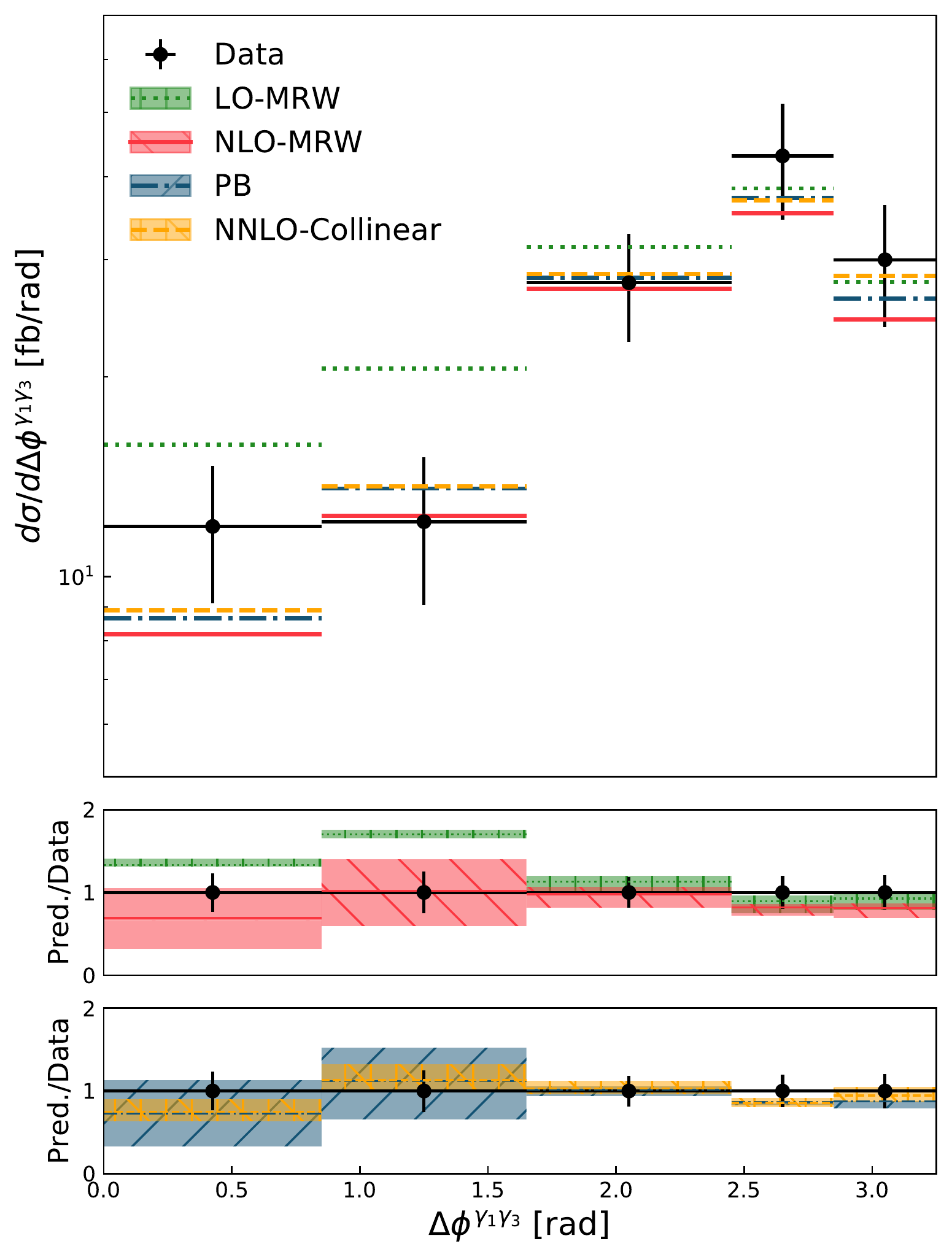}
	\includegraphics[width=8cm, height=9cm]{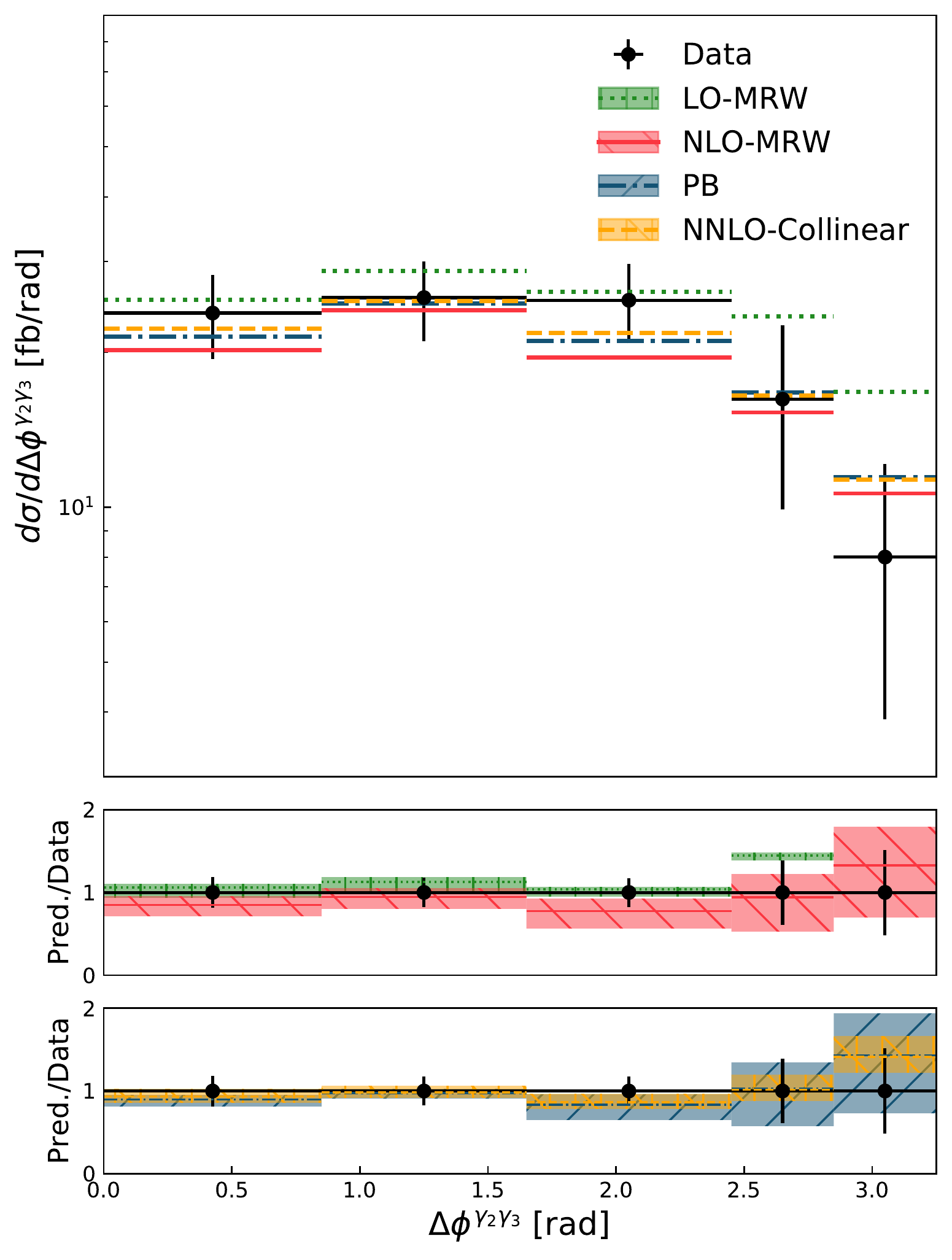}
	\caption
	{ The same as \cref{fig:6} but for difference between azimuthal angles of final state photons.	
	}
	\label{fig:7}
\end{figure}

In the \cref{fig:7}, differential cross sections predictions with respect to the difference between azimuthal angle of the three photons with the highest transverse energies, i.e. $\Delta \phi^{\gamma_1 \gamma_2}$, $\Delta \phi^{\gamma_1 \gamma_3}$ and $\Delta \phi^{\gamma_2 \gamma_3}$ are presented. It can be seen that the LO-MRW UPDFs predictions at $\Delta \phi^{\gamma_1,\gamma_2} < 2$ and $\Delta \phi^{\gamma ^{\gamma_2 \gamma_3}} > 2$ fails to describe the experimental data. While, the predictions of the NLO-MRW and PB UPDFs similar to the NNLO collinear results are in excellent agreement with the experimental data. Additionally, an important point here in these channels is that the scale uncertainties of the NLO-MRW and PB behave similar to each other and become large at small $\Delta \phi^{\gamma_1\gamma_2}$ and $\Delta \phi^{\gamma_1\gamma_3}$, while for the  $\Delta \phi^{\gamma_2\gamma_3}$ their scale uncertainties become significant at larger azimuthal angle differences.

\begin{figure}
	\includegraphics[width=8cm, height=9cm]{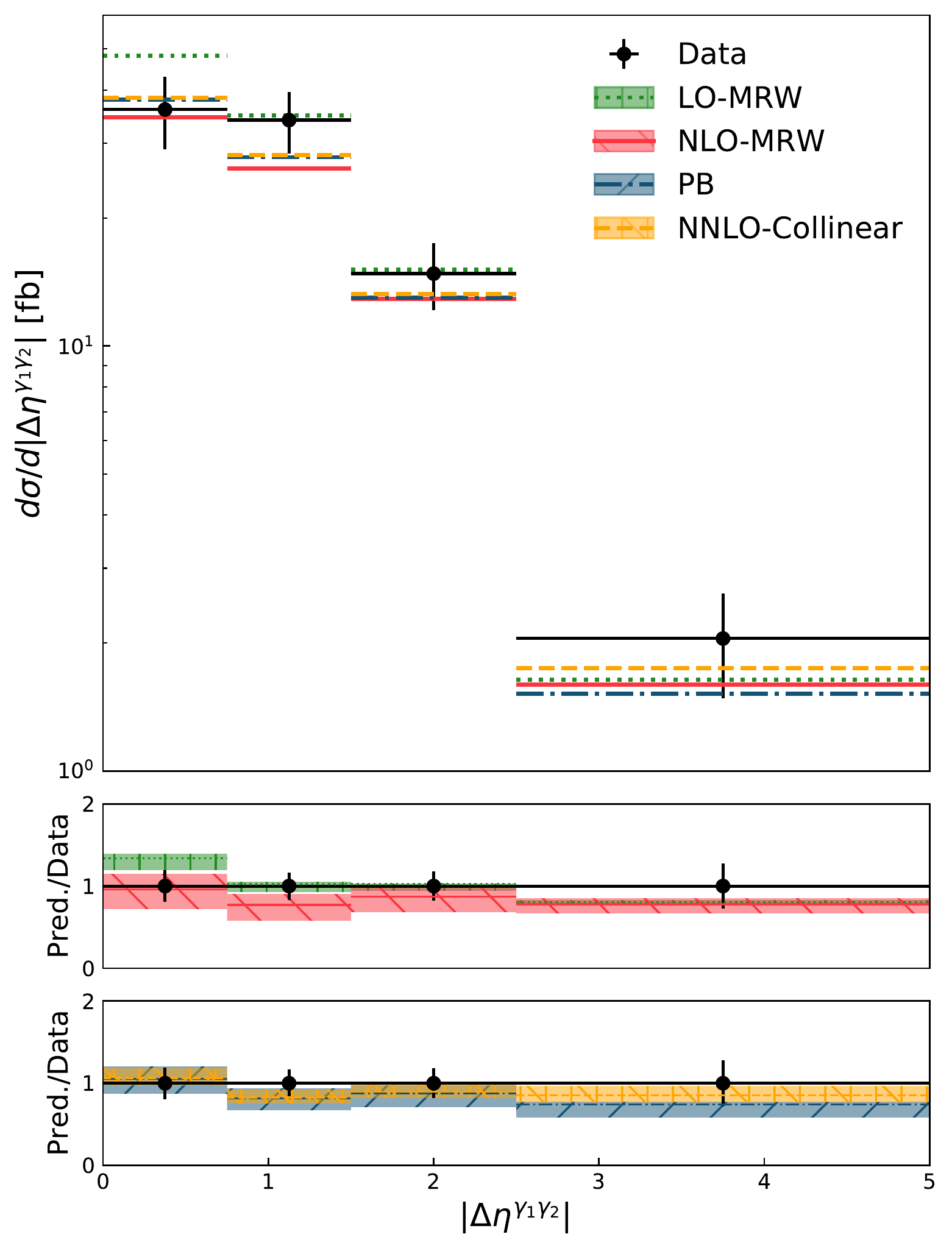}
	\includegraphics[width=8cm, height=9cm]{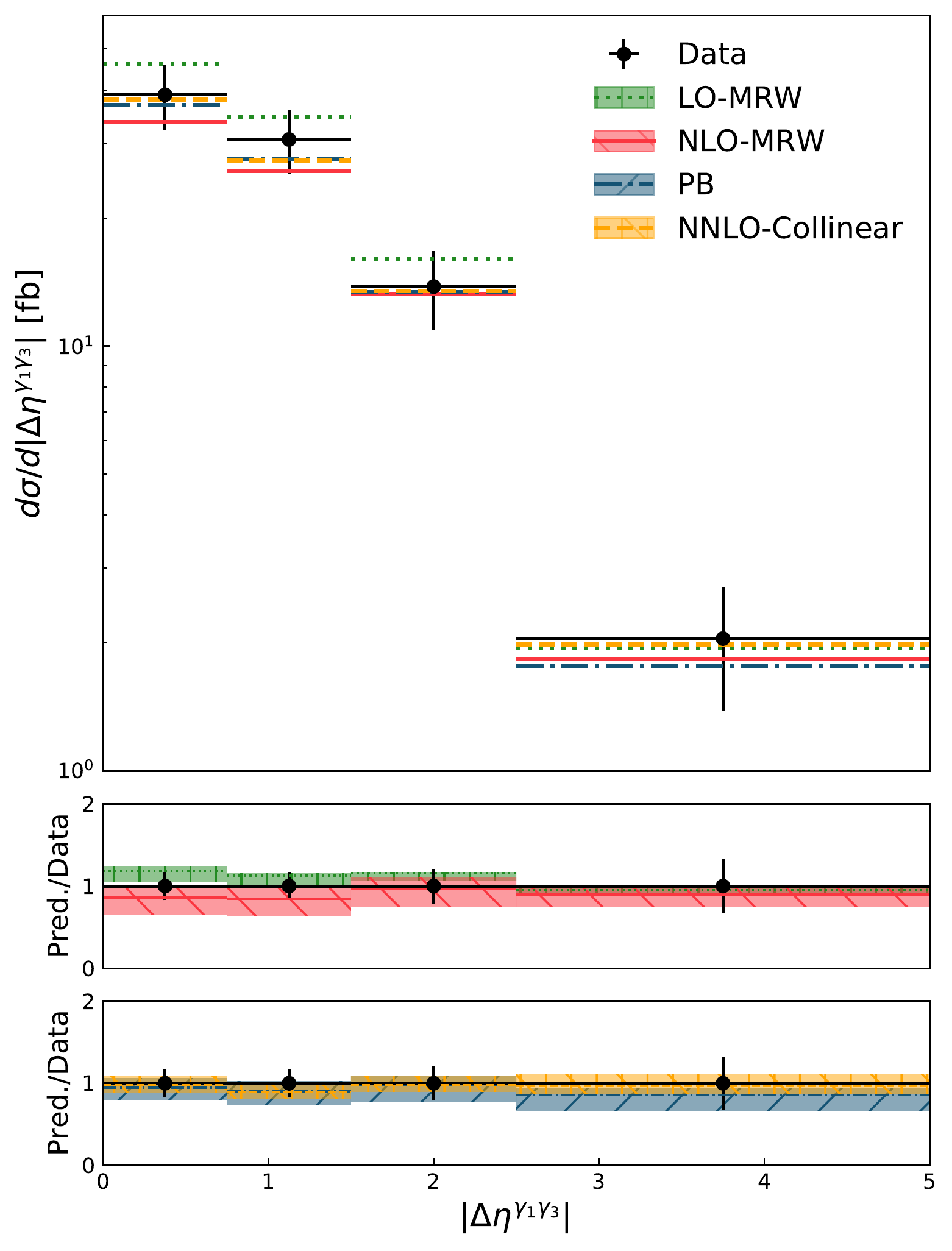}
	\includegraphics[width=8cm, height=9cm]{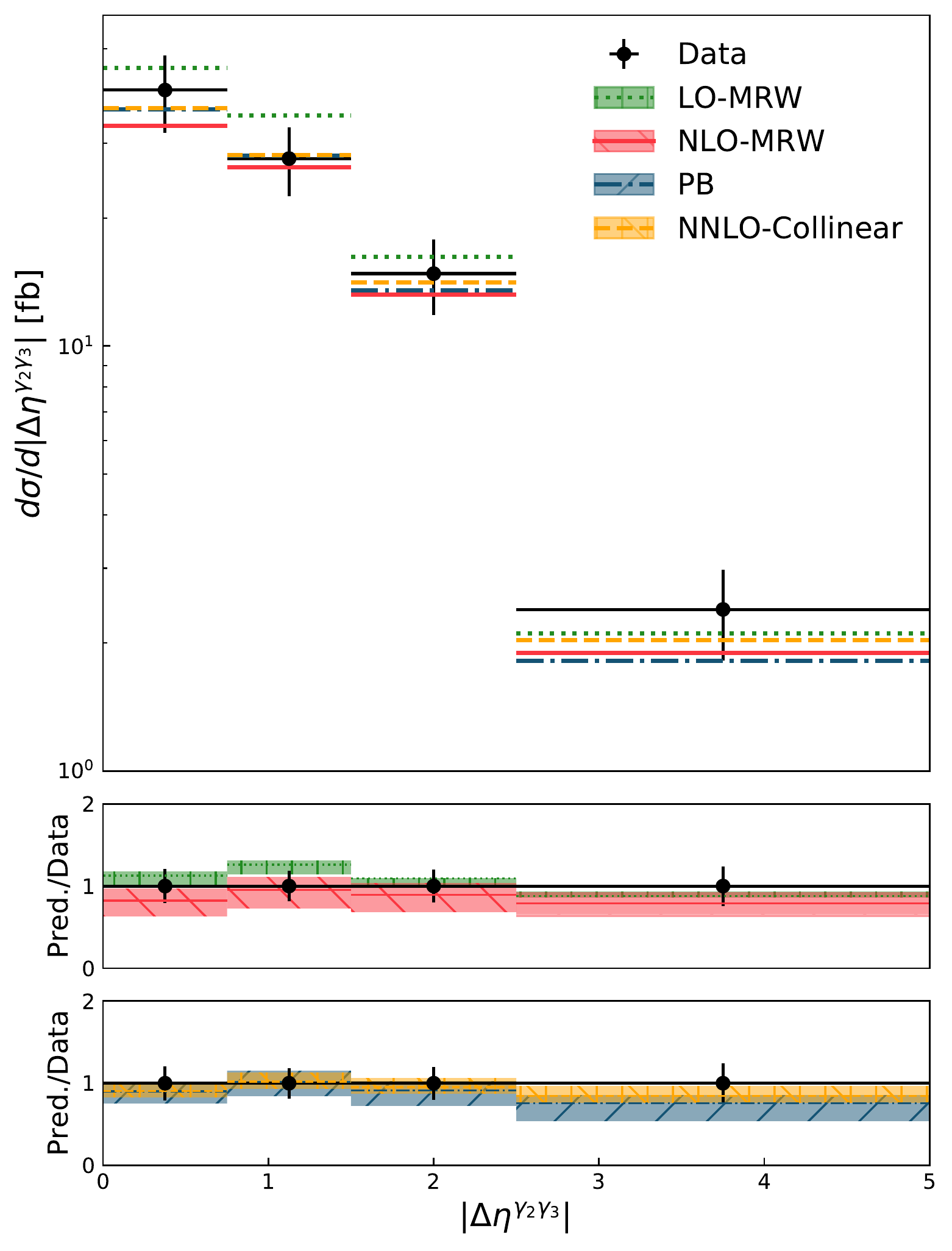}
	\caption
	{  The same as \cref{fig:6} but for absolute value of difference between pseudo-rapidities of final state photons.	
	}
	\label{fig:8}
\end{figure}

In the \cref{fig:8}, differential cross sections with respect to the difference between pseudo-rapidites of the final state photons, i.e. $|\Delta \eta ^{\gamma_i \gamma_j}|$, where $i \neq j$ and $i=1,2,3$ or $j=1,2,3$, are shown. It can be seen that when $|\Delta \eta ^{\gamma_i \gamma_j}|$ is getting smaller the LO-MRW predictions become larger and even tend to overestimate the data, while at large $|\Delta \eta ^{\gamma_i \gamma_j}|$ all models behave like to each other. 

\begin{figure}
	\includegraphics[width=8cm, height=9cm]{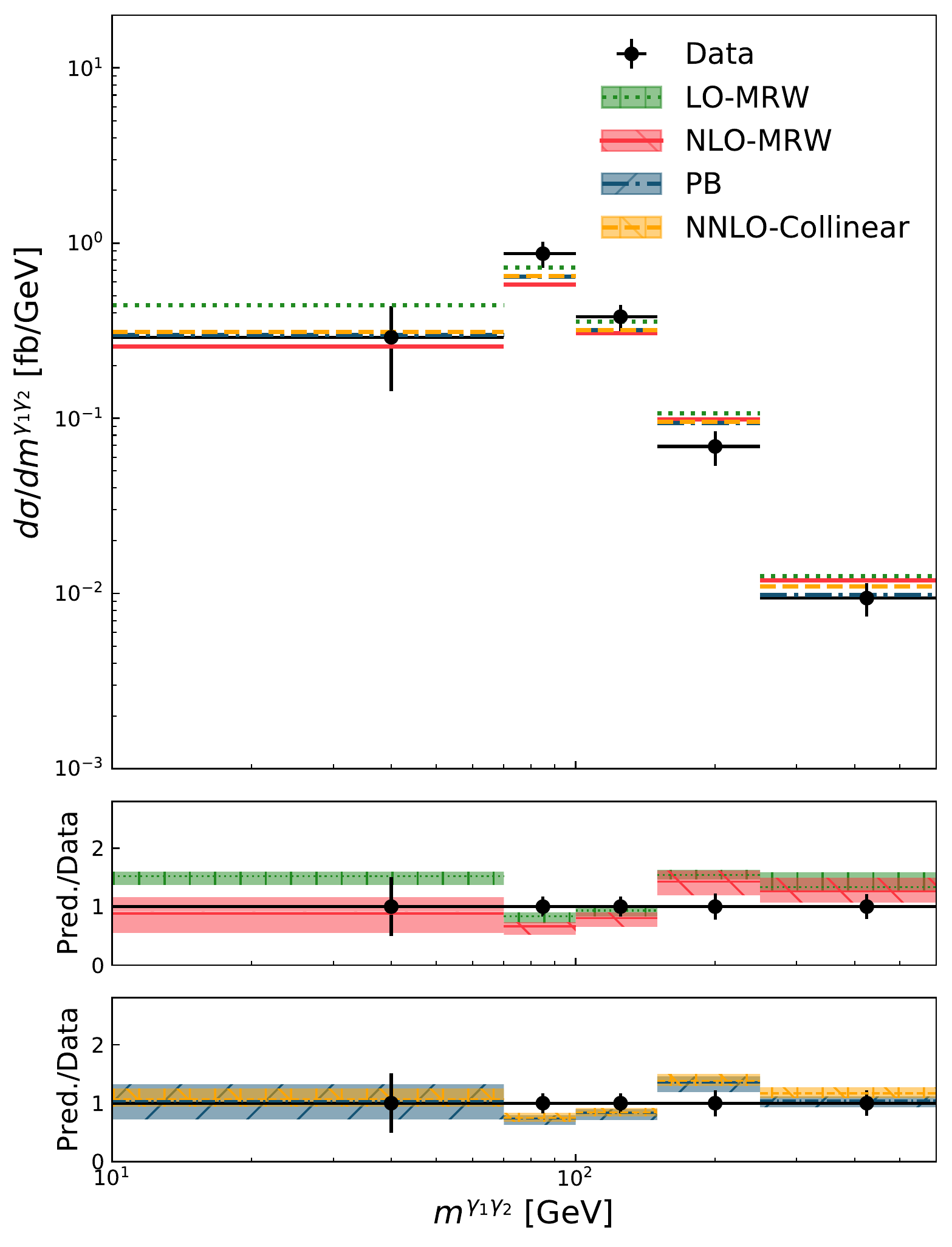}
	\includegraphics[width=8cm, height=9cm]{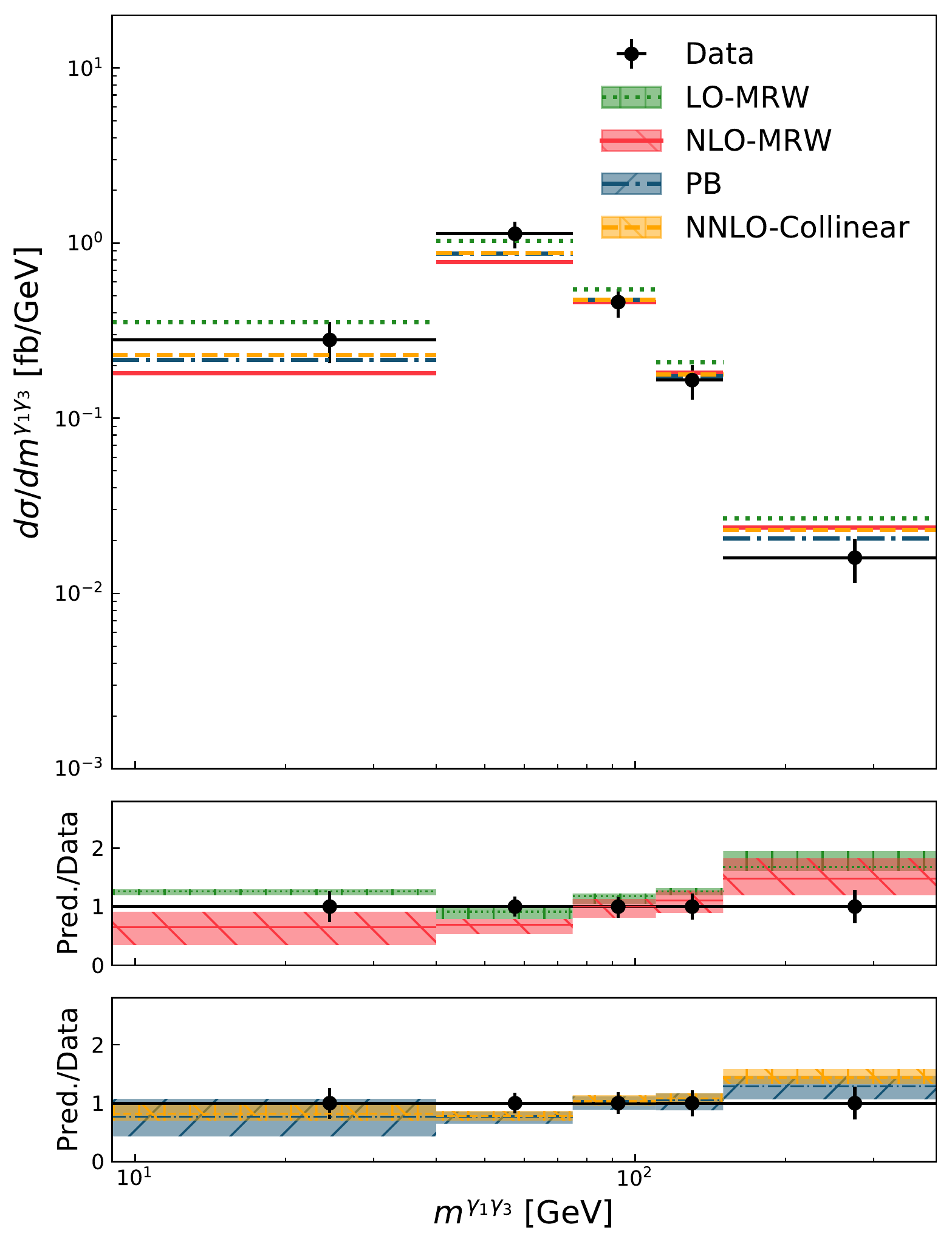}
	\includegraphics[width=8cm, height=9cm]{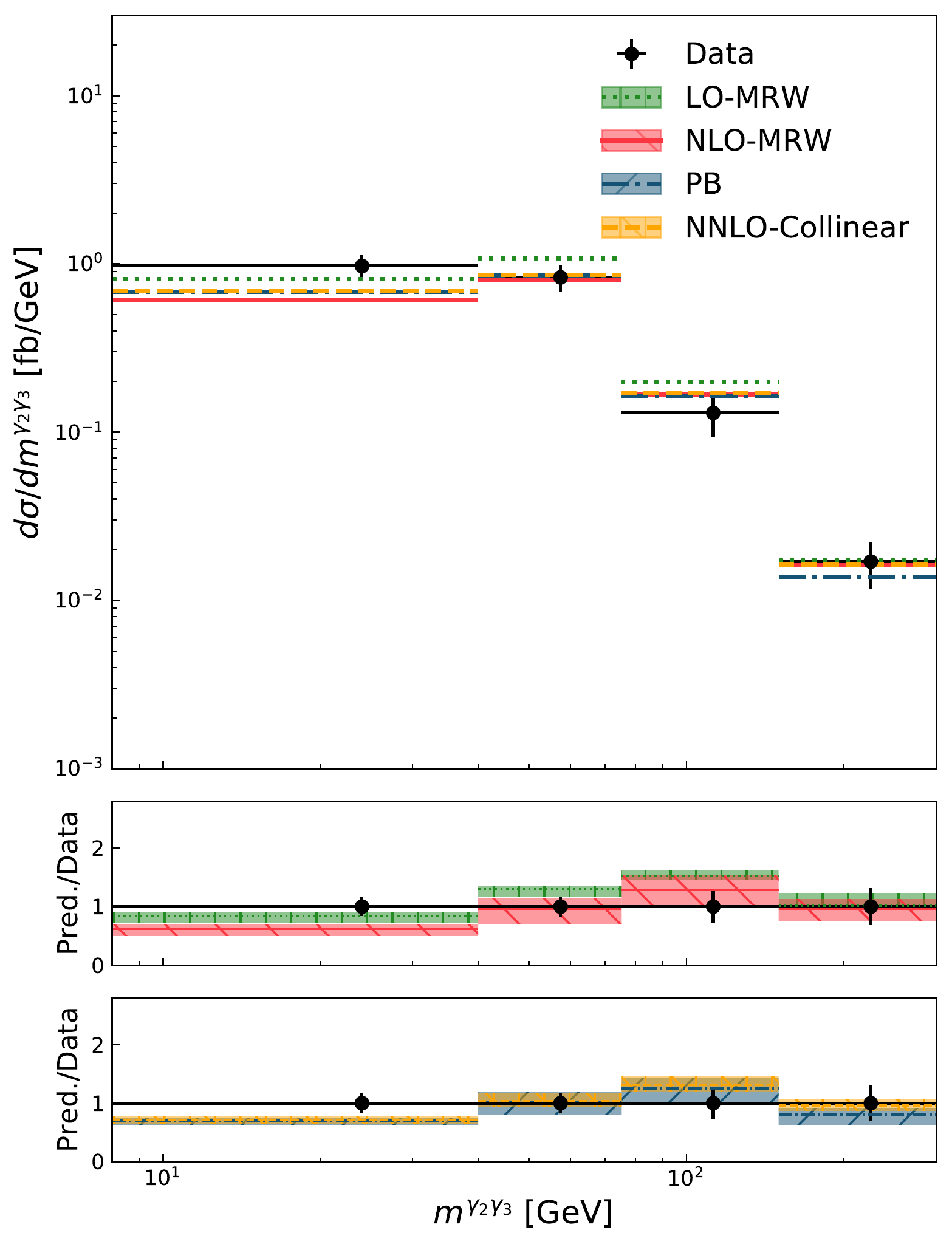}
	\includegraphics[width=8cm, height=9cm]{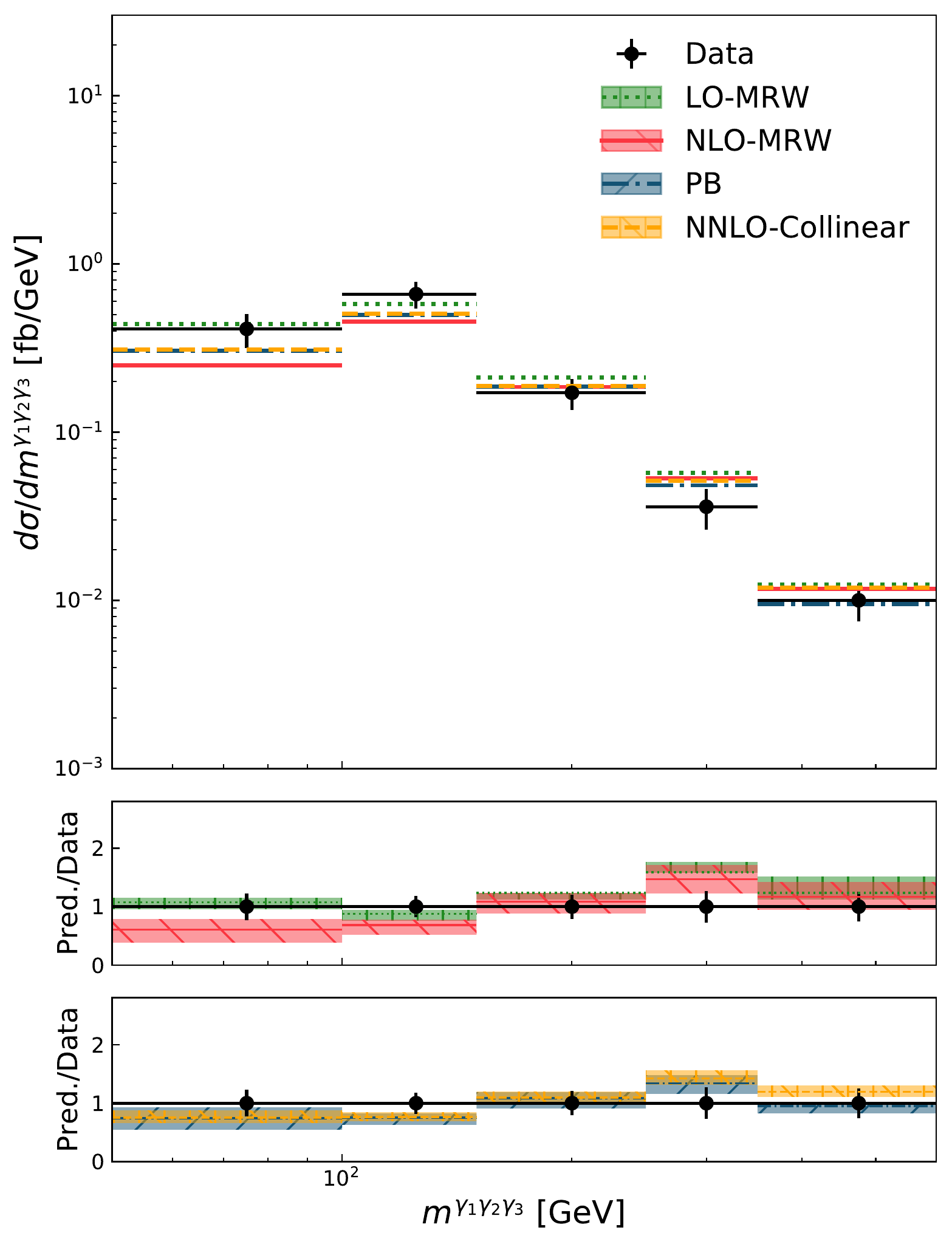}
	\caption
    {The same as \cref{fig:6} but for invariant masses of different configuration of final state photon.
	}
	\label{fig:9}
\end{figure}

Finally, in the \cref{fig:9}, differential cross section predictions with respect to invariant mass of various configurations of final state photons, i.e. $m^{\gamma_1 \gamma_2}$, $m^{\gamma_1 \gamma_3}$, $m^{\gamma_2 \gamma_3}$ and $m^{\gamma_1 \gamma_2 \gamma_3}$ are presented. It can be seen that small invariant mass is more sensitive to the transverse momentum of incoming parton and difference between UPDF models mostly shown themselves in this region. However, despite different behaviors of each UPDF models, all of them are in good agreement with the experimental data.

\section{conclusions}
\label{sec:4}
In this work, we presented the three-photon production cross section of the $8\;TeV$ ATLAS collaboration data within $k_t$-factorization framework with the LO-MRW, NLO-MRW and PB UPDF models. The calculation was done with the help of the KATIE parton level event generator and it was observed that the NLO-MRW and PB predictions are giving more similar results with respect to the LO-MRW model and also their predictions are close to the ones within the NNLO collinear . Whereas the results with the LO-MRW UPDF overshoots the data of some channels specially those with respect to the azimuthal angle difference. 

To understand better the difference between the results of UPDF models, we provided various plots in  $\log_{10}x-\log_{10}k_t^2$ space. These plots gave a detailed comparison between each UPDF models and it is showed that LO-MRW UPDFs become larger than other two UPDF models at $k_t \sim \mu$. While PB and NLO-MRW UPDFs are larger with respect to the NLO-MRW UPDF at small transverse momentum. Additionally, a similar behavior between the NLO-MRW and PB UPDFs is observed, where this similarity in case of the gluon UPDF is more striking.  

In conclusion, it was observed that the $k_t$-factorization framework, especially with a proper UPDF models such as PB and NLO-MRW, one can obtain a satisfactory description of the ATLAS three photon production data that are comparable even to the NNLO collinear results. While, it was observed in \cite{three_photons_ATLAS} that the NLO collinear is unable to give a good description of the data.

\newpage

\end{document}